\documentclass[iicol, sn-mathphys, Numbered, pdflatex]{sn-jnl}

\usepackage{xcolor, mathtools, subfig}

\usepackage{graphicx}%
\usepackage{amsmath,amssymb,amsfonts}%
\usepackage{manyfoot}%
\begin{document}

\title{Spectral analysis of spin noise in an optically spin-polarized stochastic Bloch equation driven by noisy magnetic fields}
\author[1, 2]{M. M. Kim}\email{mmkim1998@gmail.com}
\author[1, 3]{Sangkyung Lee}\email{sklee82@add.re.kr}
\affil[1]{Emerging Science and Technology Directorate, Agency for Defense Development, Republic of Korea}
\affil[2]{ORCiD: \href{https://orcid.org/0000-0002-7582-6697}{0000-0002-7582-6697}}
\affil[3]{ORCiD: \href{https://orcid.org/0000-0002-8094-3230}{0000-0002-8094-3230}}

\abstract{We provide a closed-form autocorrelation function and power spectral density (PSD) of the solution, along a prescribed probing direction, to a noisy version of an optically pumped Bloch equation wherein each component of the external magnetic field is subject to (possibly correlated) white noise. We conclude that, up to first order in the white noise covariance amplitudes, noise in the bias B-field direction does not affect the autocorrelation function. Moreover, the noise terms for the remaining two axes make different contributions to the magnetic noise-driven spin PSD; in particular, the contribution corresponding to noises perpendicular to the probing direction dominates at high frequencies. Some results concerning the second (and higher) order terms are given, and an effective Larmor frequency shift caused by anisotropic transversal B-field noises, towards the DC direction, is revealed. The analytic results are supported by Monte Carlo simulations employing the Euler-Maruyama method.}
\maketitle

\section{Introduction}\label{sec:intro}
Magnetic resonances such as nuclear magnetic resonance (NMR) and electron paramagnetic resonance (EPR) serve as basic principles for modern atomic sensors such as atom spin gyroscopes (ASGs) \cite{LLI2020, WSN2016} and optically pumped atomic magnetometers (OPAMs) \cite{BOD1961, TOP2019, JHO2012}, respectively. 
The sensitivity of an atomic sensor is determined by various noise sources such as the photon shot noise, the atomic spin projection noise (intrinsic spin noise), and the noise induced by undesired magnetic fields from the surrounding electronics (e.g., electric heaters) and the environment. Among these, photon shot noise and atomic spin projection noise set the fundamental sensitivity limit of the atomic sensor. Atomic spin projection noise can be reduced by utilizing spin-squeezed states, which are produced by quantum nondemolition measurements \cite{SMS2012, GSO2005}. Photon shot noise can be reduced by utilizing squeezed light \cite{WSO2010, TSE2021}. However, the sensitivities tend to be degraded from the fundamental limit due to magnetic field noises. Such noises from various electronics can be suppressed by employing sensor components generating low levels of magnetic field noise. For example, electric heaters for heating up atom cells can be designed to minimize the resulting magnetic interference. Such designs can be realized by double-layered polyimide films \cite{YDP2018} or by employing SOS-CMOS technology \cite{KMA2016}. On the other hand, OPAMs operated in the Earth's magnetic field cannot avoid the white-like noise from their environments, including the surrounding electronics. In ASGs and OPAMs, as the spin collections are polarized by optical pumping, the magnetic field noise-induced spin noise can dominate the intrinsic atomic spin projection noise.  For a fully polarized spin system, the atomic spin projection noise is given as $\sigma_{\text{in}} = \sqrt{F/(2 N_a)}$ where $N_a$ is the number of atoms in the collection and $F$ does not carry units of $\hbar$ \cite{RQN2013}. In contrast, the magnetic field noise-induced spin noise depends on the degree of the spin polarization. For example, we show that the magnetic field noise-induced spin noise can be written as $\sigma_{\text{ex}} \simeq \left| \gamma P \right| \sqrt{ \delta B^2_{x}+\delta B^2_{y}} / (4\sqrt{\Gamma_2}) $ given that the Larmor frequency is much larger than the transversal spin relaxation rate $\Gamma_2$, where $P=2S_z$ is the degree of polarization, $\gamma$ is the gyromagnetic ratio, and $\delta B_x$ and $\delta B_y$ are the magnetic field noise amplitudes along the $x$ and $y$ directions, respectively. In this paper, we focus on the regime where $\sigma_{\text{ex}} \gg \sigma_{\text{in}}$. The precise mathematical formulation for such a regime, including the interpretation for $\delta B_x$ and $\delta B_y$, will become clearer later after we compare the noise powers for these two types of noises in Section~\ref{sec:theory}.


Although the field noise-driven spin precession is complex, we can still analytically investigate the spin noise in the presence of white magnetic field noise, and provide some new insight into atomic sensors. In this regard, we investigate the response of a magnetic resonance system subject to external B-field white noise by analyzing the corresponding stochastic differential equation.

In this paper, we focus on the Bloch equation, which is one of the simplest phenomenological equations describing magnetic resonance \cite{WSN2016, BNI1946, SUT2004, SDI2008, LHS2016},
\begin{equation} \label{eq:bloch}
\frac {d \vec{\mathbf{S}}} {dt} = \gamma \vec{\mathbf{S}} \times \left( \vec{\mathbf{B}} +\delta \vec{\mathbf{B}} \right) - R (\vec{\mathbf{S}} - \frac{\vec{\mathbf{s}}}{2}) - \mathbf{\Gamma_{\text{re}}} \vec{\mathbf{S}},
\end{equation}
where $\vec{\mathbf S}$ is the average spin (e.g., the average on a hot $^{87}$Rb vapor cell), $\gamma$ is the gyromagnetic ratio, $R$ is the optical pumping rate, $\vec{\mathbf{s}}$ is the photon spin vector \cite{ATO1998}, and $\mathbf{\Gamma_{\text{re}}}$ is the spin relaxation matrix. In this paper, we assume that $\delta \vec{\mathbf{B}}$ is white noise, $\vec{\mathbf{B}}$ is along the z-direction, and the pumping beam is also $z$-directional with a $\sigma^{+}$ polarization, i.e., $\vec{\mathbf{s}}=\mathbf{\hat{z}}$. This equation includes an optical pumping term so that it describes a motion of \emph{polarized} spins in the presence of the white magnetic field noise. We note that the stochastic Bloch equation for unpolarized spins was previously analyzed in \cite{LHS2016} by means of the stochastic path integrals. As it turns out in this paper, in the spin-polarized case ${\mathbf S}_z = P/2$, even with a small external magnetic field noise, the spin noise is mainly determined by the fluctuations of the polarized spin driven by the magnetic field noise. A similar remark applies when we compare this paper to a related experiment on an activated crystal \cite{SSS2019}. We also note that as Eq.~\eqref{eq:bloch} focuses on spin noise driven by a noisy magnetic field, it contrasts to studies such as \cite{MEO2022}, which incorporated spontaneous fluctuations due to, e.g., collisions between atoms, or \cite{LCF2017}, which focused on noise due to atomic diffusion.

Studies tend to describe the power spectral density (PSD) of the spin noise $S_x$, arising from various noise sources, as a Lorentzian function or a sum of nearly Lorentzian functions, see: \cite{IFO1973, SHB2010, RQN2013, STO2016}. Moreover, to the best of our knowledge, no one has explicitly noted the fact that the PSD should depend on the noise direction (for cases in which such a direction is fixed); we can intuitively speculate that $x$-directional noises and $y$-directional noises lead to different PSDs of $S_x$, because we are observing the $x$-direction.

\begin{figure} [!h]

\begin{minipage}[t]{.5\linewidth}
\centering
\subfloat[]{\label{fig:highint}\includegraphics[scale=0.17]{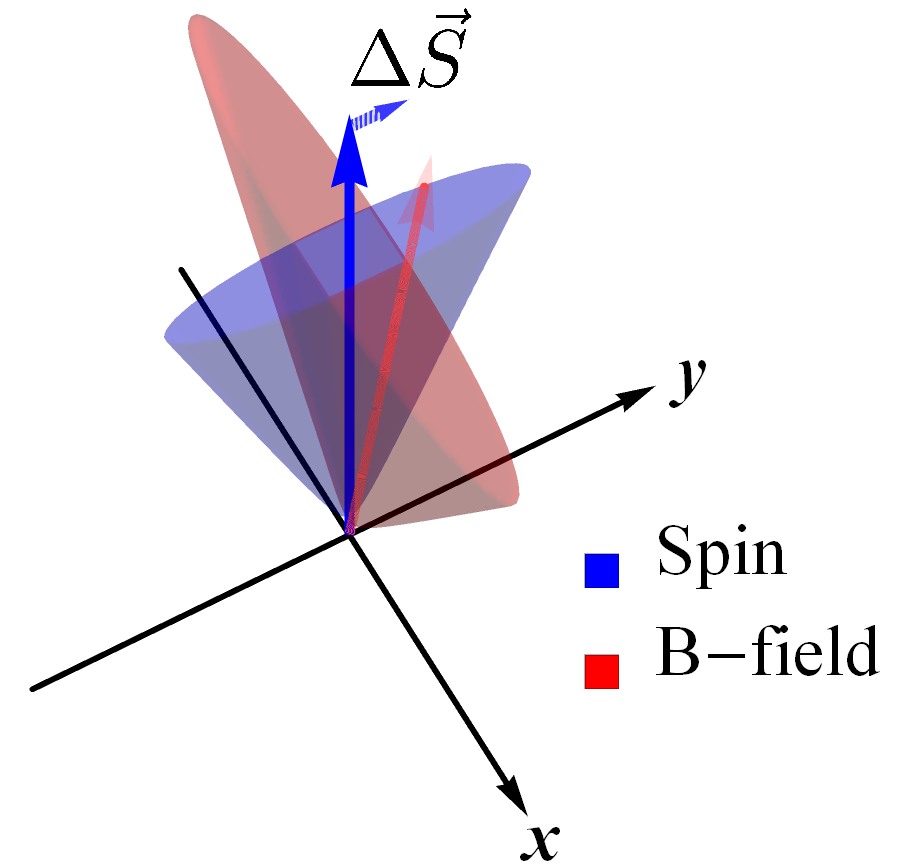}}
\end{minipage}%
\begin{minipage}[t]{.5\linewidth}
\centering
\subfloat[]{\label{fig:lowint}\includegraphics[scale=0.17]{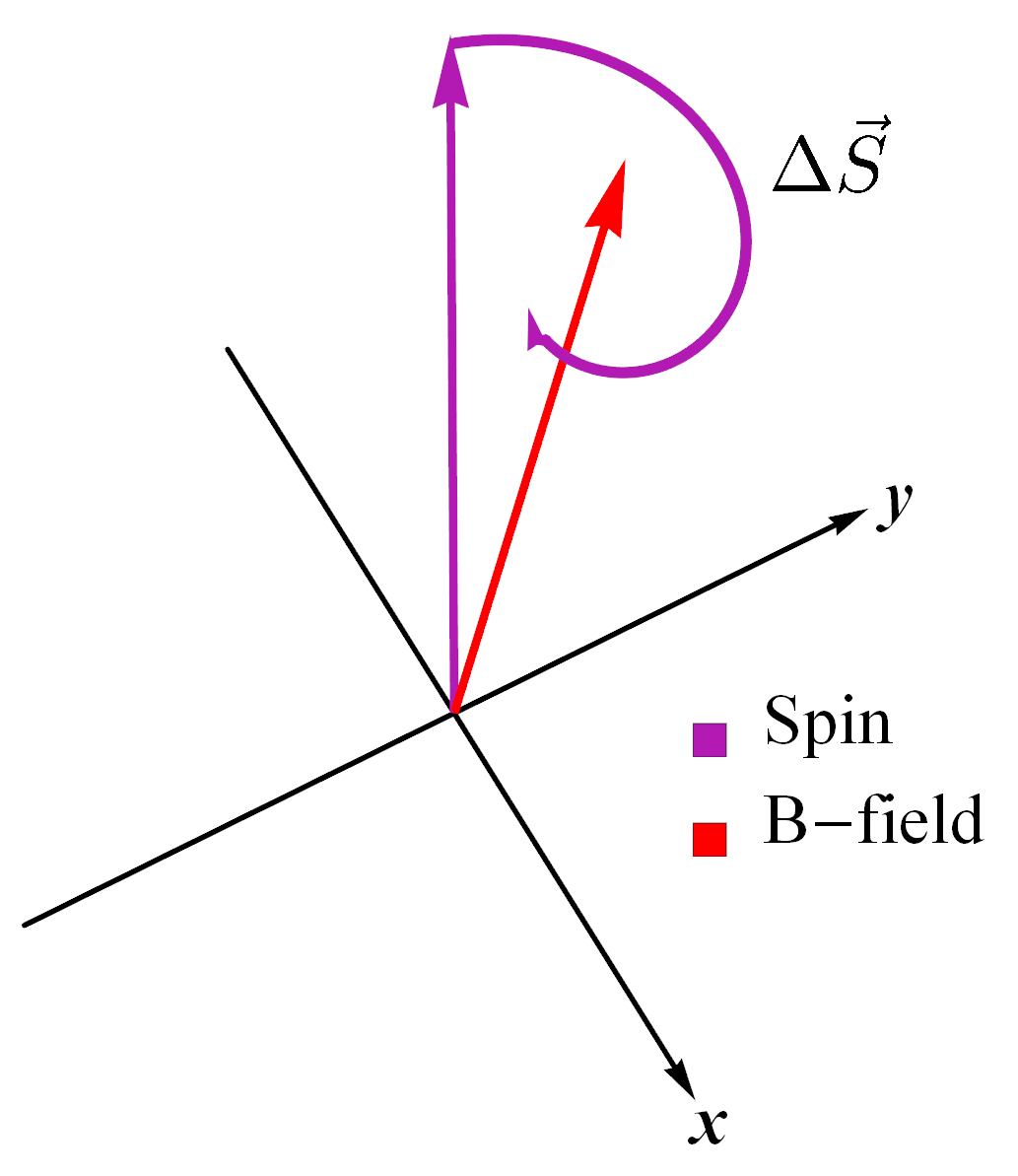}}
\end{minipage}\par\medskip
\centering
\subfloat[]{\label{fig:highlowexp}\includegraphics[width=\linewidth]{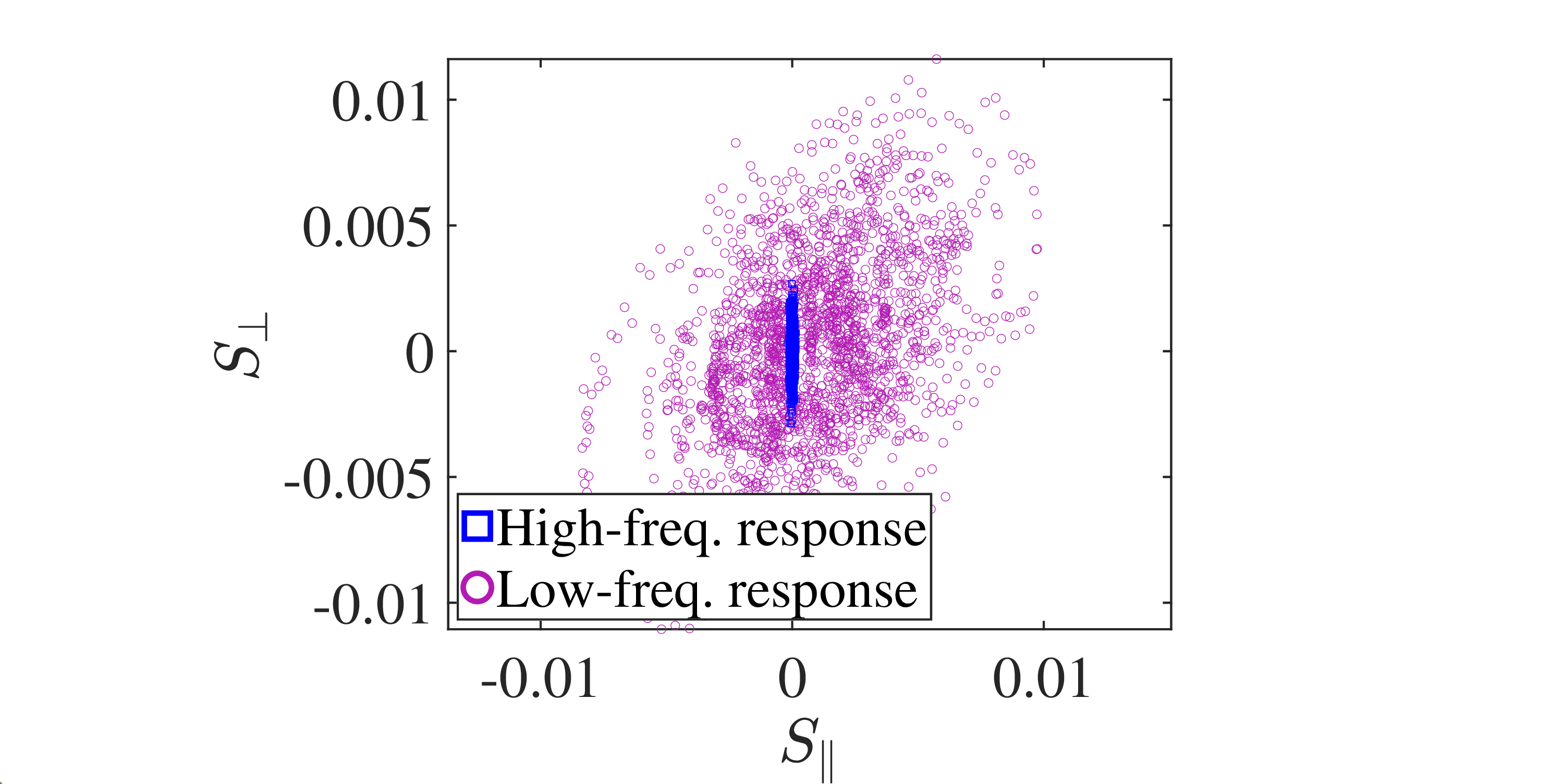}}

\caption{(a) Intuition of the system's response to the high-frequency components of an $x$-directional white noise. The precession axis (red) is blurred, leading to transitory motions of the spin vector (blue). Being transitory, these motions are mainly perpendicular to the noise direction. (b) Intuition of the response to the low-frequency components. In this case the precession axis is somewhat fixed; therefore, the spin vector can precess around a well-defined axis for a while. Therefore, spin motions parallel to the noise direction occur. (c) A sample transversal spin trajectory corresponding to a high-frequency noise and a low-frequency noise, drawn by a computer simulation. The horizontal axis $S_{\parallel}$ of the graph denotes the spin component parallel to the field noise direction, whereas the vertical axis $S_{\perp}$ corresponds to the component perpendicular to the noise. Note that $\langle S_{\parallel} S_{\perp} \rangle$ appears to be non-zero, as can be verified from Eq.~\eqref{eq:SxSy}.} 
\label{fig:main}
\end{figure}

Moreover, a close inspection of the magnetic resonance system reveals that such anisotropy would depend on the frequency range of interest. For the high-frequency range of the PSD, which will correspond to high-frequency magnetic field noises, the Larmor precession axis $\vec{\mathbf{B}}+\delta \vec{\mathbf{B}}$ is hastily altered along the field noise direction before a spin component along that direction is sufficiently created. Hence, the spin motion is mainly distributed on the axis perpendicular to the field noise axis (see Fig.~\ref{fig:highint}). In other words, spin noises parallel to the magnetic field noise are suppressed. On the other hand, when a low-frequency magnetic field noise comes in, the precession axis does not change hastily; therefore, sufficient spin precession can occur, enabling spin noise parallel to the magnetic field noise to come into the picture (see Fig.~\ref{fig:lowint}).

We note that such a frequency-dependence can be confirmed by computer simulations. Fig.~\ref{fig:highlowexp} shows a sample transversal spin noise trajectory corresponding to two types of transversal magnetic field noises: a high-frequency noise (blue) and a low-frequency noise (purple). Details of the simulation are found in Appendix A.

In this paper, we validate the above-mentioned intuitions by analytically solving Eq.~\eqref{eq:bloch} up to first order in the B-field noise covariance amplitudes (these amplitudes will be denoted as $\Gamma_{\alpha \beta}$). Moreover, we show that the PSD of $S_x$ cannot be written as a sum of nearly Lorentzian functions. Finally, we give some second-order corrections to the PSD. We believe that such higher order information cannot be obtained if we approximate Eq.~\eqref{eq:bloch} by a Langevin equation: assuming $\delta B_z(t) = 0$ for all $t$ and approximating $S_z$ to be a constant transforms the equation into a Langevin form. Such higher order terms exhibit some qualitatively different features from their first-order counterparts: the higher order lineshapes are much more complicated and they exhibit longitudinal-transversal coupling which is absent in the first-order terms.

\section{Theory} \label{sec:theory}
We consider a situation in which $\vec{\mathbf{B}} = B \mathbf{\hat z} + \begin{bmatrix}  \delta B_x (t)& \delta B_y(t)& \delta B_z(t) \end{bmatrix}^{\text{T}}$, where $\delta B_\alpha(t)$ are mean-zero noise terms with an approximately white power spectrum. We note that the average spin $\vec{\mathbf S}$ of the collection experiences a spatially homogeneous magnetic field fluctuation as we are implicitly assuming the size of the atomic vapor cell is much smaller than the spatial scale of the B-field fluctuations, in contrast to the inhomogeneous fluctuations which lead to additional spin relaxations \cite{CRO1988}. We assume a diagonal relaxation matrix with $\Gamma_2'$, $\Gamma_2'$, and $\Gamma_1'$ as its three consecutive diagonal entries, where $\Gamma_2'$ and $\Gamma_1'$ are the transversal and longitudinal spin relaxation rates excluding the contributions of optical pumping. Note that the relaxation matrix is transversally ($x$, $y$) isotropic, although it is not fully isotropic as $\Gamma_1'$ may differ from $\Gamma_2'$ \cite{BNI1946}. We moreover assume $\vec{\mathbf{s}}=\mathbf{\hat{z}}$, i.e., a $\sigma^{+}$ pumping beam. In this case, the Bloch equation can be written in the following matrix form.

\begin{equation} \label{eq:blochmatrixform}
\begin{split}
\frac{d\vec{\mathbf{S}}}{dt} &= \left(\begin{bmatrix} -\Gamma_2'-R & \gamma B &0 \\ -\gamma B & -\Gamma_2'-R & 0\\ 0 & 0&-\Gamma_1' - R\end{bmatrix}\right. \\
&+ \left. \gamma \begin{bmatrix} 0 & \delta B_z  & -\delta B_y \\ -\delta B_z & 0 & \delta B_x\\\delta B_y & -\delta B_x&0\end{bmatrix} \right) \vec{\mathbf{S}} + \frac{R}{2} \mathbf{ \hat z}
\end{split}
\end{equation}

From now on, we denote $\Gamma_1'+R$ and $\Gamma_2'+R$ as $\Gamma_1$ and $\Gamma_2$, respectively. We assume $\gamma B \neq 0$ and $\Gamma_1$, $\Gamma_2$ to be positive. Denoting the right-hand side of Eq.~\eqref{eq:blochmatrixform} as $(\mathbf{A_1}+\mathbf{A_2} (t)) \vec{\mathbf{S}} + \vec{\mathbf{c}}$, and by defining a coordinate change as $\mathbf{\widetilde {S} }(t) \coloneqq \text{exp} \left( -t \begin{bmatrix} \mathbf{A_1} & \vec{\mathbf{c}} \\ \mathbf{0} & 0 \end{bmatrix} \right)  \begin{bmatrix} \vec{\mathbf{S}} \\ 1\end{bmatrix}$, the equation is transformed into the following homogeneous form.
\begin{equation}
\frac{d\mathbf{\widetilde {S}}}{dt} = \exp \left( -t \begin{bmatrix}\mathbf{ A_1 }& \vec{ \mathbf c} \\ \mathbf 0 & 0 \end{bmatrix} \right)   \begin{bmatrix} \mathbf{A_2} & \mathbf 0 \\ \mathbf 0 & 0 \end{bmatrix}  \text{exp} \left( t \begin{bmatrix} \mathbf{A_1} & \vec{\mathbf c} \\ \mathbf 0 & 0 \end{bmatrix} \right)   \mathbf{\widetilde S} (t)
\end{equation}
Hereafter, we denote the right-hand-side of the above equation as $\mathbf{\widetilde A} (t) \mathbf{\widetilde S} (t)$.

We have the following second-order approximation for the autocorrelation function of $\widetilde {S}$, where $\mathbf{1}$ denotes the 4-by-4 identity matrix and h.o.t. means higher order terms.

\begin{equation} \label{eq:transformedcoordinates}
\begin{split}
\langle \mathbf{\widetilde {S}} (t_0+\tau) \mathbf{\widetilde {S} }(t_0) ^\text{T} \rangle &= \left( \vphantom{\int_{t_0}^{t_0+\tau} } \mathbf{1} + \int_{t_0}^{t_0+\tau}\!\!\!\int_{t_0} ^t  \langle \mathbf{\widetilde{A}} (t) \mathbf{\widetilde{A}}(t') \rangle dt'dt \right.\\[1ex]
&\left.+\text{(h.o.t.)}\vphantom{\int_{t_0}^{t_0+\tau}}\right) \langle \mathbf{\widetilde {S}} (t_0) \mathbf{\widetilde {S}} (t_0)  ^\text{T} \rangle
\end{split}
\end{equation}

From now on, the analysis proceeds in two steps. The first step is to calculate the time-evolution operator in the parenthesis in Eq.~\eqref{eq:transformedcoordinates}, which is achieved in Eq.~\eqref{eq:originalcoordinates}. The second step is to calculate the initial value $\langle \mathbf{\widetilde {S}} (t_0) \mathbf{\widetilde {S}} (t_0)  ^\text{T} \rangle$ when the system reaches steady state (i.e., $t_0\to\infty$); the result of this step is summarized in Eq.~\eqref{eq:equilvalues}.
Also, we note that an additional analysis involving the higher-order terms in Eq.~\eqref{eq:transformedcoordinates} will be discussed later.

Calculating the time-evolution operator and returning to the original coordinates, we have the following expression for the autocorrelation function of $S_x$:
\begin{strip}
\begin{equation} \label{eq:originalcoordinates}
\begin{split}
&\langle S_x (t_0+\tau) S_x(t_0) \rangle \\
=&\text{ } e^{-\Gamma_2\tau} \left( \cos (\tau \omega) +\frac{\gamma^2 (\Gamma_{xx} - \Gamma_{yy})}{4 \omega } \sin (\tau \omega) -\frac{\gamma^2  (\Gamma_{xx} + \Gamma_{yy}+2\Gamma_{zz})}{4} \tau \cos(\tau\omega) \right) \langle S_x (t_0)^2 \rangle \\
+&\text{ } e^{-\Gamma_2\tau} \left( \left(1+\frac{\gamma^2 \Gamma_{xy}}{2 \omega}\right)\sin(\tau\omega)  -\frac{\gamma^2 (\Gamma_{xx} + \Gamma_{yy}+2\Gamma_{zz})}4 \tau \sin(\tau\omega) \right)   \langle S_x (t_0) S_y(t_0) \rangle \\
+&\text{ } \mathfrak{C} \langle S_x (t_0) S_z(t_0) \rangle +\mathfrak{D} \langle S_x(t_0) \rangle,
\end{split}
\end{equation}
\end{strip}
where we denoted the Larmor frequency $\gamma B$ as $\omega$. Here, we assumed that the noises are white with their covariances given by delta functions as $\langle \delta B_{\alpha} (t) \delta B_{\beta} (t') \rangle = \Gamma_{\alpha \beta} \delta(t-t')$ for symmetric coefficients $\Gamma_{\alpha \beta}$; that is, the correlation time of the B-field noises are much shorter than any relevant time scales of the system. $\mathfrak{C}$ and $\mathfrak{D}$ are complicated correlation terms that vanish when the coefficients $\Gamma_{\alpha \beta}$ for $\alpha \neq \beta$ are zero (the full expressions are given in Appendix C).

Approximation \eqref{eq:originalcoordinates} for the autocorrelation function of $S_x$ is a priori valid only for short evolution times $\tau$ with $\tau \mathbf{\widetilde A} \ll 1$ (or more precisely, $\int_{t_0}^{t_0 + \tau}  dt \mathbf{\widetilde A}(t) \ll 1$). Meanwhile, we argue that equation \eqref{eq:originalcoordinates} also holds for $\tau$ large enough so that $\Gamma_{1} \tau \text{, }\Gamma_{2} \tau \gg 1$. First, the right-hand-side of the equation clearly vanishes whenever $\Gamma_{1} \tau \text{, }\Gamma_{2} \tau \gg 1$, at least up to first order in $\Gamma_{\alpha \beta}$. Also, the left-hand side can be approximated by $\langle S_x (t_0+ \tau) \rangle \langle S_x (t_0) \rangle$ for such large $\tau$ because physical intuition tells us that $S_x(t_0)$ and $S_x(t_0+\tau)$ will be statistically uncorrelated; any deviation from the expectation at time $t_0$ will be exponentially damped out in time, thereby not affecting the spin at time $t_0+\tau$. Therefore, both sides of the equation vanish up to first order in $\Gamma_{\alpha \beta}$, assuming the system is at its steady-state. We emphasize that this argument hinges on the change of coordinates into $\mathbf{\widetilde {S}}$, which resulted in the damping exponential $e^{-\Gamma_2\tau}$ terms shown in the right-hand-side of Eq.~\eqref{eq:originalcoordinates}.

\begin{figure}[!h]
\centering
\includegraphics[width=0.5\textwidth, trim = 0cm 20cm 0cm 0cm, clip]{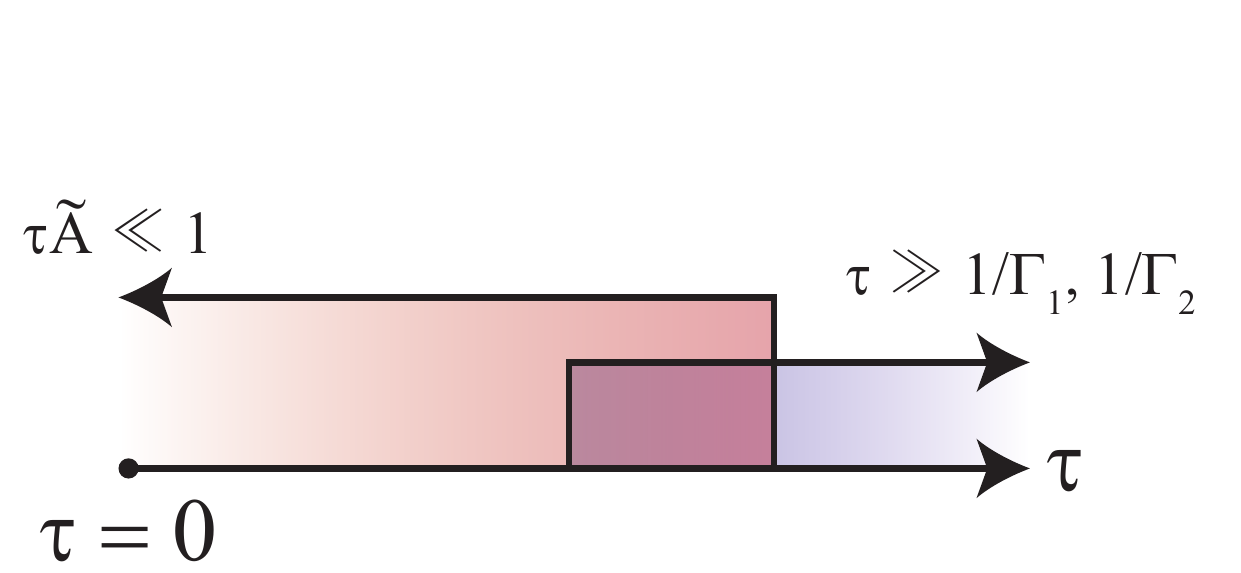} 
\caption{Eq.~\eqref{eq:originalcoordinates} is valid for all $\tau \geq 0$ for sufficiently small noises.}
\label{fig:approx}
\end{figure}

To sum up, Eq.~\eqref{eq:originalcoordinates} is valid for values of $\tau$ satisfying $\tau \mathbf{\widetilde A} \ll 1$ or $\Gamma_{1} \tau \text{, }\Gamma_{2} \tau \gg 1$. As a result, it will be valid for all $\tau \geq 0$ as long as these two regions overlap: see Fig.~\ref{fig:approx}. Such an overlap happens when the noise is small enough, since in this case the region $\tau \mathbf{\widetilde A} \ll 1$ depicted in Fig.~\ref{fig:approx} will be shifted towards $\tau \to \infty$.

In any case, equation \eqref{eq:transformedcoordinates} is exact if we consider all the higher order terms \cite{RLS1995}. Later in this section, the effect of the first non-zero higher order term is considered; if the resulting correction terms are much smaller than the first-order terms in $\Gamma_{\alpha \beta}$, we can, in retrospect, conclude that the noise was small enough for the approximation in \eqref{eq:transformedcoordinates} to be valid.

So far, we did not assume that the system is in its steady state. Now, we determine the relevant steady state values $\langle S_x (t_0) ^2 \rangle$, $\langle S_x (t_0) S_y(t_0) \rangle$, $\langle S_x (t_0) S_z(t_0) \rangle$, and $\langle S_x (t_0) \rangle$ after all the irrelevant transients are damped out (i.e., $t_0\to\infty$). This can be achieved by enlarging the main vector variable further to include the second-order terms $\mathbf{\mathcal{S}}(t)\coloneqq \begin{bmatrix}S_x(t) S_x(t) &S_x(t) S_y(t) &\cdots &S_z(t) S_z(t) \end{bmatrix} ^{\text{T}}$ as in the following equation.

\begin{equation} \label{eq:secondorderterms}
\begin{split}
\frac{d} {dt} \begin{bmatrix} \mathbf{\mathcal{S}} \\ \vec{\mathbf S} \\1  \end{bmatrix} &= \left(\begin{bmatrix} \mathbf{A_1} \otimes \mathbf{1} + \mathbf{1} \otimes \mathbf{A_1} & \vec{\mathbf c} \otimes \mathbf{1} + \mathbf{1} \otimes \vec{\mathbf c} &\mathbf 0 \\ \mathbf 0 &\mathbf{A_1} &\vec{\mathbf c}\\\mathbf 0&\mathbf 0&0\end{bmatrix}\right.\\
&+ \left.\begin{bmatrix} \mathbf{A_2} \otimes\mathbf{1} + \mathbf{1} \otimes \mathbf{A_2} &\mathbf 0&\mathbf 0\\\mathbf 0&\mathbf{A_2}&\mathbf 0\\\mathbf 0&\mathbf 0&0 \end{bmatrix} \right) \begin{bmatrix} \mathcal{S} \\ \vec {\mathbf S} \\1 \end{bmatrix}
\end{split}
\end{equation}

Here, $\mathbf{1}$ denotes the 3-by-3 identity matrix and $\otimes$ denotes the Kronecker product. Denoting the terms in the parenthesis as $\mathcal{A}_1 + \mathcal{A}_2$, and defining a coordinate change $\widetilde{\mathcal{S}} (t) \coloneqq e^{-t \mathcal{A}_1}  \begin{bmatrix} \mathcal{S}(t) & \vec{\mathbf S}(t) &1  \end{bmatrix}^{\text{T}}$, we conclude that the second-order approximation for $\langle \widetilde{\mathcal{S}} (t)  \rangle$ similar to equation \eqref{eq:transformedcoordinates} coincides with the first-order approximate solution to the following differential equation \eqref{eq:approxODE}. Although such an equivalence is a priori valid only for sufficiently short time evolutions, this limitation is superficial because \eqref{eq:approxODE} is independent of the initial time $t_0$.

\begin{subequations}
\begin{alignat}{2} 
\frac{d} {dt} \langle \widetilde{\mathcal{S}} (t)  \rangle =&\text{ }e^{-t \mathcal{A}_1} \mathbf{\widetilde {\Gamma}}  e^{t\mathcal{A}_1} \langle \widetilde{\mathcal{S}} (t)  \rangle\\
\mathbf{\widetilde {\Gamma}} \coloneqq& \int_{t_0} ^t  \langle \mathcal{A}_2(t) \mathcal{A}_2 (t') \rangle dt'
\end{alignat} \label{eq:approxODE}
\end{subequations}
Here, $\mathbf{\widetilde {\Gamma}}$ is a 13-by-13 constant matrix that is independent of $t>t_0$. It is a matrix that has linear combinations of $\gamma^2 \Gamma_{\alpha \beta}$ as its components (the full expression is given in Appendix C).

In conclusion, up to first order in $\Gamma_{\alpha \beta}$ (i.e., up to second order in the noise amplitudes),  we have the following equation in the original coordinates.
\begin{equation} 
\frac{d} {dt} \left\langle  \begin{bmatrix} \mathcal{S} \\\vec {\mathbf S} \\1 \end{bmatrix}   \right\rangle =(\mathcal{A}_1+ \mathbf{\widetilde {\Gamma}} )  \left\langle \begin{bmatrix} \mathcal{S} \\ \vec{\mathbf S} \\1 \end{bmatrix} \right\rangle
\smallskip
\end{equation}

Incidentally, every eigenvalue of the 12-by-12 submatrix of $\mathcal{A}_1+ \mathbf{\widetilde {\Gamma}}$ at the top left corner has a negative real part when $\Gamma_{\alpha \beta}=0$; this mathematically proves that the expectations $\langle \mathcal{S} \rangle$ and  $\langle \vec{ \mathbf S} \rangle$ will indeed converge to an equilibrium, at least for sufficiently small $\Gamma_{\alpha \beta}$ values.

These steady state values can be obtained by solving a 12-by-12 linear equation; some are provided in the following equations, written up to first order in $\Gamma_{\alpha \beta}$. We denoted the value of the time variable as $\infty$ to emphasize that Eq.~\eqref{eq:equilvalues} represents the steady state values.
\begin{subequations}\label{eq:equilvalues}
\begin{align}
\begin{split} \label{eq:SxSx}
\langle S_x(\infty) S_x(\infty) \rangle &=\frac{\gamma^2 R^2}{16\Gamma_1^2 \Gamma_2 \left( \omega^2 + \Gamma_2 ^2 \right)}\\
\left( \omega^2 \Gamma_{xx}\right.&\left. -2 \Gamma_2 \omega \Gamma_{xy} + (\omega^2 +2\Gamma_2^2) \Gamma_{yy} \right)
\end{split}\\
\langle S_x(\infty) S_y(\infty) \rangle &= \frac{\gamma^2 R^2  \left(  \omega\Gamma_{xx} -2\Gamma_2 \Gamma_{xy}- \omega\Gamma_{yy} \right)}{16\Gamma_1^2  \left( \omega^2+\Gamma_2^2 \right)}   \label{eq:SxSy} \\
\langle S_x(\infty)S_z(\infty) \rangle & = \frac{\gamma^2 R^2 \left( \omega \Gamma_{yz}+\Gamma_2 \Gamma_{zx}\right)}{8 \Gamma_1^2 \left( \omega^2 + \Gamma_2 ^2 \right)} \label{eq:SxSz} \\
\langle S_x(\infty) \rangle &=\frac{\gamma^2 R \left( \omega \Gamma_{yz}+\Gamma_2 \Gamma_{zx}\right)}{4 \Gamma_1 \left( \omega^2 + \Gamma_2 ^2 \right)} \label{eq:Sxequil} \\
\langle S_z(\infty) \rangle &= \frac{R}{2 \Gamma_1} - \frac{\gamma^2 R} {4 \Gamma_1 ^2} \left( \Gamma_{xx}+\Gamma_{yy} \right)  \label{eq:Szequil} 
\end{align}
\end{subequations}
Note that the $xy$-product given in Eq.~\eqref{eq:SxSy} vanishes when $\Gamma_{xx}=\Gamma_{yy}$ and $\Gamma_{xy}=0$, as expected due to symmetry. Moreover, the steady state value for $S_z$ has a first-order correction term, which cannot be easily justified by the Langevin approach; the Langevin approach, in order to be justified, requires $S_z$ to be approximated by its a priori equilibrium value $R/2 \Gamma_1$. Finally, we note that \eqref{eq:SxSz} and \eqref{eq:Sxequil} vanish whenever the correlation coefficients $\Gamma_{yz}$ and $\Gamma_{zx}$ are zero.

Plugging the relevant steady state values into \eqref{eq:originalcoordinates}, we finally obtain the autocorrelation function at steady state. The autocorrelation function, the one-sided PSD, and the total power of $S_x(t)$ are given in the following equations, which are written up to first order in $\Gamma_{\alpha \beta}$. It turns out that there is no first-order effect of nonzero $\Gamma_{zz}$, $\Gamma_{yz}$, or $\Gamma_{zx}$. That is, the system under consideration is insusceptible to longitudinal noises.
\begin{strip}
\begin{subequations} \label{eq:mainresults}
\begin{align}
\begin{split} \label{eq:autocorr}
 \langle S_x(t_0+\tau) S_x(t_0) \rangle  &=  \mathcal{C} \left( \omega^2 \Gamma_{xx}-2  \Gamma_2\omega \Gamma_{xy}+ (2\Gamma_2^2 +\omega^2)\Gamma_{yy}\right) \cos(\omega \tau)\\
&+ \mathcal{C} \Gamma_2 (\omega\Gamma_{xx}-2\Gamma_2\Gamma_{xy} -\omega\Gamma_{yy}) \sin(\omega \tau)\text{ }\text{ }\text{ }\text{ }\text{ for }\tau \geq 0\text{, at steady state. }\text{ } \text{ }\text{ }\text{ }\text{ }\text{ }\text{ }\text{ }\text{ }
\end{split}\\
\begin{split} \label{eq:psd}
\text{PSD}_{S_x} (f) &= 4 \int_0 ^\infty \langle S_x(t_0+\tau) S_x(t_0) \rangle \cos (2 \pi f \tau) d \tau
=\frac{\gamma^2 R^2 \left(\omega^2 \Gamma_{xx} -2\omega\Gamma_2 \Gamma_{xy}+ (\Gamma_2^2 +\Omega^2 )\Gamma_{yy} \right)}{2 \Gamma_1^2 \left(( \Omega - \omega)^2 + \Gamma_2^2 \right) \left((\Omega+\omega )^2 + \Gamma_2^2 \right)}
\end{split}\\
\begin{split} \label{eq:totalpower}
\langle S_x ^2 \rangle = \int_0 ^\infty  \text{PSD}_{S_x} (f) df &=\frac{\gamma^2 R^2 \left(\omega^2 \Gamma_{xx} -2\Gamma_2 \omega \Gamma_{xy}+ (\omega^2+2\Gamma_2^2 )\Gamma_{yy} \right)}{16\Gamma_1^2 \Gamma_2 \left( \omega^2 +\Gamma_2^2  \right)}
\end{split}
\end{align}
\end{subequations}
\end{strip}
In Eq.~\eqref{eq:mainresults}, $\mathcal{C} \coloneqq \frac{\gamma^2 R^2 e^{-\Gamma_2 \tau}}{16\Gamma_1^2 \Gamma_2 \left(\Gamma_2^2 + \omega^2 \right)}$ and $\Omega \coloneqq 2 \pi f$. Note that \eqref{eq:totalpower} coincides with \eqref{eq:SxSx} as expected. Eq.~\eqref{eq:psd} follows from the Wiener-Khinchin theorem. From Eq.~\eqref{eq:totalpower}, we can derive the following condition for the intrinsic spin noise to be dominated by the field noise-driven spin noise.
\begin{equation}
\frac{\gamma^2 R^2 \Gamma_{xx \text{ or } yy} }{\Gamma_1^2 \Gamma_2 } \gg \frac{1}{N_a}
\label{eq:fieldnoise_dominant_condition}
\end{equation}
Here, $N_a$ denotes the number of atoms in the collection \cite{RQN2013}. In case of a cubic $87$-rubidium cell of side length $5$ mm and temperature $80$ $^{\circ}\text{C}$, and assuming that the pumping and the relaxation parameters are given as in Section~\ref{sec:mc}, this will hold for $\Gamma_{xx \text{ or } yy} \geq  10^{-5} \text{ nT}^2\cdot \text{ms}$ which corresponds to $\delta B_{\alpha} \geq 100~\text{fT}/\sqrt{\text{Hz}}$ where $\delta B_{\alpha}^2 \coloneqq \Gamma_{\alpha \alpha}$. Note that by applying $P = 2 S_z \approx R/\Gamma_1  $ and $\delta B_{\alpha}^2 = \Gamma_{\alpha \alpha}$ to Eq.~\eqref{eq:totalpower} and then by taking square roots, we arrive at the expression $\sigma_{\text{ex}} \simeq \left| \gamma P \right| \sqrt{ \delta B^2_{x}+\delta B^2_{y}} / (4\sqrt{\Gamma_2})$ which was previously noted in the Introduction \ref{sec:intro}. Such a slight change in notation explicitly reveals the role of the degree $P$ of the spin polarization.

In equation \eqref{eq:totalpower}, the contribution of $\Gamma_{yy}$ to the total power is larger than that of $\Gamma_{xx}$. Our result agrees with physical intuition; for example, an $x$-directional noise tilts the Larmor precession axis slightly along the $x$-axis, so that the precessions projected on the $xy$-plane will be somewhat compressed along the $x$-axis, leading to smaller $S_x$ magnitude compared to the case of a $y$-directional noise input. Moreover, as the Larmor frequency $\omega$ grows larger, discrimination between the $x$ and $y$ directions will be gradually blurred, so that the discrepancy becomes negligible for $\omega \gg \Gamma_2$.

Note that we have an additional ``high-pass'' term $ \Omega^2 \Gamma_{yy}$ in the numerator of equation \eqref{eq:psd}. This term dominates the $\Gamma_{xx}$ term when $\omega \ll \Omega$. Therefore, $y$-directional noise is somewhat more ``high-passed'' compared to $x$-directional noise. This agrees with the intuitive predictions presented in the Introduction. In case $\omega \ll \Omega$, we imagine $\vec{\mathbf S}$ as being along the $z$-direction, wherein some transitory displacements $\delta \vec{\mathbf S}$ appear due to a ``blurring'' of the precession axis (Fig.~\ref{fig:highint}). On the other hand, in case $\omega \gg \Omega$, the precession axis $\vec{\mathbf B}+\delta\vec{\mathbf B}$ can now be imagined to be permanently tilted (Fig.~\ref{fig:lowint}). In this case there is enough time for the precession to evolve so that $x$-directional magnetic noise can now contribute to the PSD of $S_x$, as long as $\Gamma_2$ is not too large to wipe out the precession. (Such a wipe-out could happen when the B-field is sufficiently small, in the order of a few tens of nanoteslas in the case of $^{87}$Rb.) Therefore, in this case, the relative strength between $\Gamma_2$ and $\omega$ will be an issue; indeed, it is the case as can be seen in equation \eqref{eq:psd}.

Also, we note that there is no first-order contribution of $\Gamma_{zz}$ to the PSD. This is clear, since without any $x$ or $y$-directional perturbation to the bias field, the spin will be fixed along the $z$-direction. However, if some transversal noise enters into the system, small precessions will occur, therefore enabling the $z$-directional noise to affect the whole picture. However, this will be a second-order effect involving the products $\Gamma_{xx}\Gamma_{zz}$ or $\Gamma_{yy}\Gamma_{zz}$.

We note that a directional transversal noise $\delta\vec{\mathbf B}(t)= \delta B(t) \left( \cos\theta  \mathbf{\hat x}+\sin\theta \mathbf{\hat y} \right)$ with $\langle \delta{B}(t) \delta{B}(t')\rangle= \Gamma_{\perp} \delta(t-t')$ results in the following noise power, which is clearly non-negative.
\begin{equation} \label{eq:turnaround}
\langle S_x ^2 \rangle = \frac{ \gamma^2 R^2 \Gamma_{\perp} \left(\Gamma_2^2+\omega^2 -\Gamma_2^2 \cos(2\theta)-\Gamma_2 \omega \sin(2\theta) \right)}{16\Gamma_1^2 \Gamma_2 \left( \omega^2 +\Gamma_2^2  \right)}
\end{equation}
From the equation, it is clear that the noise-induced signal power is minimized at $\theta = {\pi}/4 - \arg{\left( \omega + i\Gamma_2 \right)}/2$.

Finally, we note a non-Lorentzian behavior of the first-order PSD given in Eq.~\eqref{eq:psd}. First, note that the two-sided PSD, i.e. $\text{PSD}_{S_x} (f)/2$, equals the following expression for all $\Omega$:
\begin{equation}
\begin{split}
\frac{\gamma^2 R^2}{16 \Gamma_1 ^2 \left(\omega^2 + \Gamma_2^2 \right)} &\left( \frac{2\omega^2 \Gamma_{xx} + 2 \Gamma_2 ^2 \Gamma_{yy}- \omega(\Gamma_{xx} - \Gamma_{yy})\Omega }{(\Omega - \omega)^2 
+ \Gamma_2^2}\right.\\
&\left.+\frac{2\omega^2 \Gamma_{xx} + 2 \Gamma_2 ^2 \Gamma_{yy}+ \omega(\Gamma_{xx} - \Gamma_{yy})  \Omega}{(\Omega + \omega)^2 + \Gamma_2^2} \right)
\end{split}
\end{equation}
where we assumed $\Gamma_{xy}=0$ for the sake of clarity. Each fraction in the parentheses has a $\Omega$-proportional term in its numerator, deviating it from a Lorentzian function. This additional term, at $\Omega \simeq \pm \omega$, is always comparable to the Lorentzian term in case of an $x$-directional noise; in case of a $y$-directional noise, it may even dominate the Lorentzian term, provided that $\Gamma_2 \ll \omega$, leading to a non-Lorentzian behavior of the PSD. However, if we focus on the near-resonance region so that we could replace the $\Omega$s in the numerators by $\omega$, the equation reduces to a sum of Lorentzians as
\begin{equation} \label{eq:Lorentzianapprox}
\frac{\gamma^2 R^2}{16 \Gamma_1 ^2 } \left( \frac{\Gamma_{xx}+\Gamma_{yy} }{(\Omega - \omega)^2 
+ \Gamma_2^2}+\frac{3 \Gamma_{xx} - \Gamma_{yy}}{(\Omega + \omega)^2 + \Gamma_2^2} \right)
\end{equation}
in cases where $\Gamma_2 \ll \omega$ and $\Gamma_{xx} \simeq \Gamma_{yy}$.

Now, we point out some alterations to the theory required for the second-order analysis. For this purpose, we will assume that third-order moments of the external noise vanish, and fourth-order moments satisfy the following equation; these assumptions were motivated by Wick's theorem:
\begin{equation}
\begin{split}
&\langle \delta B_x (t_1)  \delta B_x (t_2)  \delta B_x (t_3)  \delta B_x (t_4) \rangle = \Gamma_{xx} ^2 \delta(t_1-t_2)\delta(t_3-t_4) \\
&+ \Gamma_{xx} ^2 \delta(t_1-t_3)\delta(t_2-t_4) +\Gamma_{xx} ^2 \delta(t_1-t_4)\delta(t_2-t_3).
\end{split}
\end{equation}
Although the above equation is written for $\delta B_x$, we assume the analogous equations for any fourfold products of the noise terms. Such assumptions are required to compute some higher order terms mentioned in Eq.~\eqref{eq:transformedcoordinates}. In particular, our second-order analysis will require a quadruple integral term in equation \eqref{eq:transformedcoordinates}:
\begin{equation}
 \int_{t_0}^{t_0+\tau}\!\!\! \int_{t_0} ^{t_1} \!\!\!\int_{t_0} ^{t_2} \!\!\!\int_{t_0} ^{t_3}  \langle \mathbf{\widetilde{A}} (t_1) \mathbf{\widetilde{A}}(t_2)\mathbf{\widetilde{A}}(t_3)\mathbf{\widetilde{A}}(t_4) \rangle dt_4 dt_3 dt_2 dt_1
\end{equation}
whereas its triple-integral counterpart vanishes.

First, we point out that:
\begin{equation}
\begin{split}
\int_\mathcal{R} F(t_1,t_2,t_3,t_4) \delta(t_1 - t_2) \delta(t_3 - t_4) d^4 \mathbf{t} \\= \frac{1}{4} \int_{t_0} ^{t_0+\tau}\!\!\! \int_{t_0}^{t_1} F(t_1,t_1,t_3,t_3) dt_3 dt_1,
\end{split}
\end{equation}
where $\mathcal{R}$ denotes the region defined by ${t_0\leq t_4 \leq t_3 \leq t_2 \leq t_1 \leq t_0+\tau}$, while similar expressions with $\delta(t_1 -t_3)\delta(t_2-t_4)$ or $\delta(t_1 -t_4)\delta(t_2-t_3)$ terms vanish; this follows by approximating $\delta$ by a sequence of smooth functions and estimating the volume of the support of the integrand. (The number $4$ in the right-hand-side will be replaced by powers of two for the analogous higher-order equations.)
As a result, the first non-zero higher order term in equation \eqref{eq:transformedcoordinates} is simplified as the following expression:

\begin{equation} \label{eq:secondordertermDK}
\frac {\gamma^4}{4} \int_{t_0}^{t_0+\tau} \!\!\! \int_{t_0}^{t_1} e^{-t_1 \mathcal{K}} \mathcal{D} e^{ (t_1-t_3) \mathcal{K} } \mathcal{D} e^{ t_3 \mathcal{K}} dt_3 dt_1,
\end{equation}
where $\mathcal{K}=\begin{bmatrix}\mathbf{ A_1 }& \vec{\mathbf c} \\ \mathbf 0 & 0 \end{bmatrix}$ and $\mathcal{D}$ reduces, for the case in which $\Gamma_{xy}=\Gamma_{yz}=\Gamma_{zx}=0$, to the 4-by-4 diagonal matrix with its consecutive diagonal entries given by $-\Gamma_{yy}-\Gamma_{zz}$, $-\Gamma_{zz}-\Gamma_{xx}$, $-\Gamma_{xx}-\Gamma_{yy}$, and $0$ (the general form is given in Appendix C). A similar inspection shows that Eq.~\eqref{eq:approxODE} is already correct up to second order. After this, the analysis proceeds verbatim; however, one should check that Eq.~\eqref{eq:originalcoordinates} is satisfied up to second order whenever $\Gamma_1 \tau \text{, } \Gamma_2 \tau \gg 1$. This can be verified by utilizing the formulae \eqref{eq:Sxequil} and \eqref{eq:frakD}.

We can observe that the second-order coefficients for the PSD corresponding to $\Gamma_{xx}\Gamma_{zz}$, $\Gamma_{yy}\Gamma_{zz}$, and $\Gamma_{xx}\Gamma_{yy}$ are indeed non-zero (details in Appendix C), whereas the coefficient corresponding to $\Gamma_{zz}^2$ is zero.
Moreover, the coefficients for $\Gamma_{xx}^2$ and $\Gamma_{yy}^2$ are both negative at resonance. For completeness, we inform the reader that the correction terms corresponding to $\Gamma_{xx}^2$ and $\Gamma_{yy}^2$ are given as $-  \left(\omega^2 \Gamma_{xx}^2 + (\Gamma_2^2 +\Omega^2 )\Gamma_{yy}^2 \right) \mathcal{F}$
where $\mathcal{F}$ is given as follows.
\begin{equation}
\begin{split}
\mathcal{F}&= \gamma^4 R^2 \Big(\Omega^4 - \Omega^2 \left(2 \omega ^2-\Gamma_1 \Gamma_2 - 2 \Gamma_2^2 \right) \\
& + \left( \omega^2+\Gamma_2^2 \right) \left( \omega ^2+\Gamma_1 \Gamma_2 +  \Gamma_2^2 \right) \! \Big) \\
&\left.\middle/ 2 \Gamma_1^3 \left( (\Omega- \omega)^2+\Gamma_2^2 \right)^2  \left( (\Omega+\omega)^2+\Gamma_2^2 \right)^2  \right.
\end{split}
\end{equation}

Since we gave the second-order terms corresponding to $\Gamma_{xx}^2$ and $\Gamma_{yy}^2$, and the first-order terms corresponding to $\Gamma_{xx}$ and $\Gamma_{yy}$, we can now quantify the smallness of noise required for the first-order approximation to be valid: it is when the second-order terms are much smaller than the corresponding first-order terms for all $\Omega$ in the range of interest.

Now we investigate on the higher-order terms of the spin autocorrelation function in the presence of \emph{Gaussian} white noises, albeit the corresponding assumptions for the higher-order moments will be increasingly difficult to verify in practice. Although it is computationally difficult to obtain the exact analytic formulae for higher order terms, we can proceed to obtain the eigenvalues of an ODE representation for the spin autocorrelation function, which was previously approximated by a Dyson series in Eq.~\eqref{eq:transformedcoordinates}. This enables one to extract various effective relaxations and frequency shift for the spin variable.

The time evolution of the autocorrelation function can be summarized by the ODE
\begin{equation}
\begin{split}
&\frac{d}{dt} \left< \begin{bmatrix} \vec{\mathbf{S}}(t) \\ 1\end{bmatrix} \begin{bmatrix} \vec{\mathbf{S}}(t_0)^{\text{T}} & 1\end{bmatrix} \right>\\
=&  \left(\begin{bmatrix} \mathbf{ A_1 }  & \vec{\mathbf{c}} \\0&0\end{bmatrix} + \frac{\gamma^2} 2 \mathcal{D} \right)\left<
\begin{bmatrix} \vec{\mathbf{S}}(t) \\ 1\end{bmatrix} \begin{bmatrix} \vec{\mathbf{S}}(t_0)^{\text{T}} & 1\end{bmatrix}\right>,
\end{split}
\end{equation} although we still need to calculate the steady state values as before (as in Eq.~\eqref{eq:equilvalues}). Such an effective renormalization of the Bloch operator by a noise term $\mathcal D$ was previously noted in \cite{LHS2016}, wherein the effects of longitudinal noise component were investigated. Here, we may extend the results in \cite{LHS2016} by computing the eigenvalues of the renormalized Bloch operator. In the case of $\Gamma_{yz}=\Gamma_{zx}=0$, we have the following eigenvalues:
\begin{equation} \label{eq:eigenvals}
\begin{split}
-\Gamma_1 - \frac {\gamma^2} 2 (\Gamma_{xx} + \Gamma_{yy}) \text{ and}\\
-\Gamma_2 - \frac {\gamma^2} 4 (\Gamma_{xx} + \Gamma_{yy}+2 \Gamma_{zz})\pm \mathrm{i} \mkern 1mu \omega_d
\end{split}
\end{equation}
where
\begin{equation} \label{eq:omegasubd} \omega_d = \sqrt{\omega^2 - \frac{\gamma^4} {16} \left( \left(\Gamma_{xx}-\Gamma_{yy} \right)^2 + 4 \Gamma_{xy} ^2 \right)}.
\end{equation}
The first (or latter) formula in Eq.~\eqref{eq:eigenvals} describes the longitudinal (or transversal) relaxation. They show that the longitudinal relaxation rate is broadened by $\Gamma_{xx}$ and $\Gamma_{yy}$, whereas the transversal relaxation rate is broadened by any $\Gamma_{\alpha \alpha}$. For the unpolarized spin case, a similar formula was derived in Ref.~\cite{LHS2016}.

\begin{figure}[!h]
\centering
\includegraphics[width=0.45\textwidth, trim = 4cm 0cm 0cm 0cm, clip]{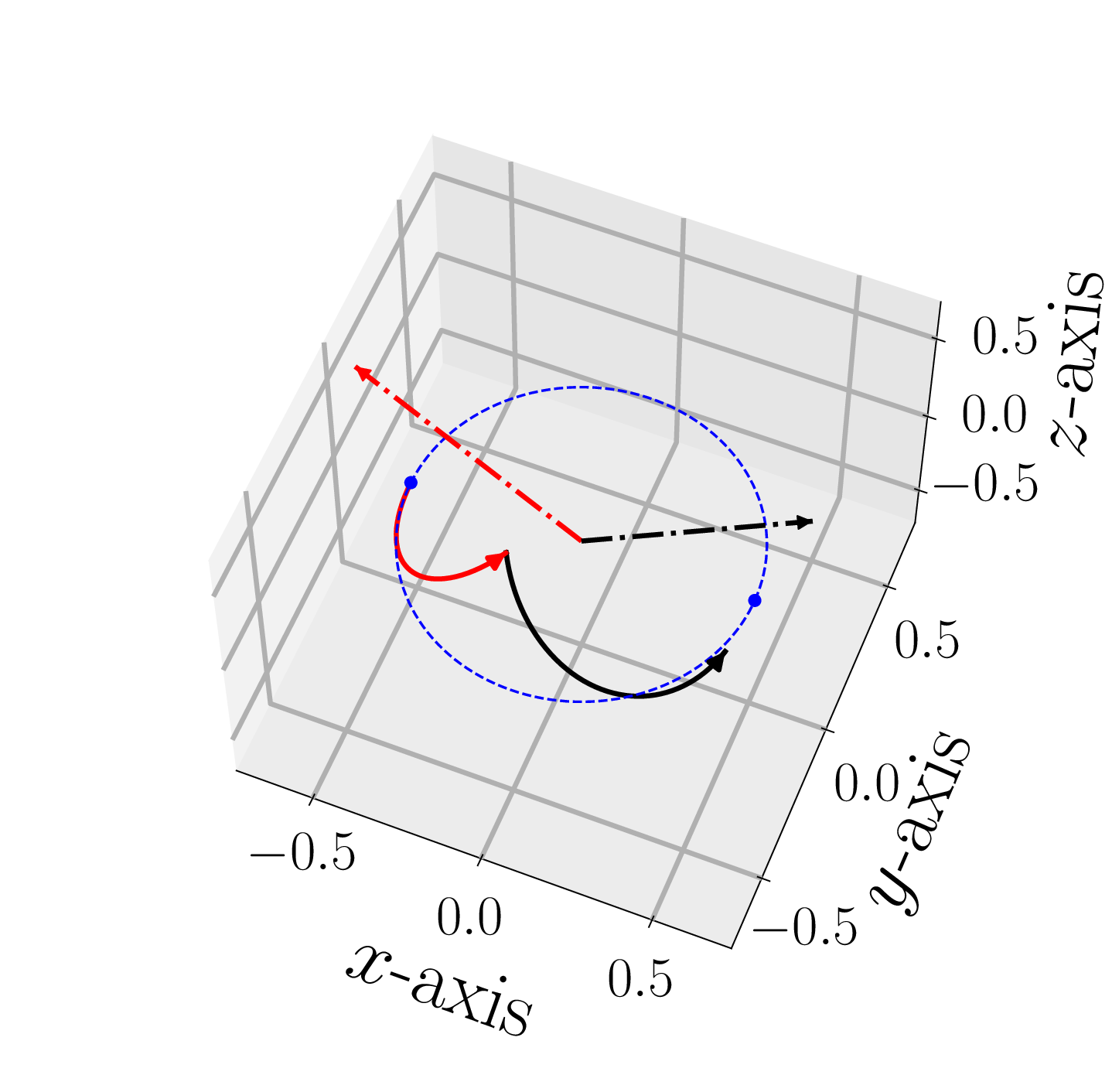}
\caption{The initial spin vector $\vec{\mathbf{S}}_{0} = (-1/2, 0, 0)$ is consecutively rotated around two different B-field axes $\vec{\mathbf B}_{\text{tot}}= \vec{\mathbf B}+\delta \vec{\mathbf{B}}(t)$. The rotations are assumed to be counterclockwise. Here, $\vec{\mathbf B} = (0,0,10)$, and $\delta \vec{\mathbf{B}}(t)$ suddenly changes from $(-10,0,0)$ to $(10,0,0)$ (arb. unit). The first rotation (solid red) points upward, whereas the second rotation (solid black) points downward. The dashed blue curve corresponds to the $xy$-planar circle of radius $1/2$ centered at the origin. The dash-dotted segments represent the axes of rotations. Blue dots correspond to $(\pm 1/2,0,0)$. Each rotation corresponds to a rotation angle of $90^{\circ}$, corrected for the instantaneous increase of Larmor frequency. As the blue dots indicate, the over precession angle is less than $180^{\circ}$. That is, the longitudinal wobbling of the spin vector trajectory rendered the overall precession angle to appear smaller.}
\label{fig:wobbling}
\end{figure}

Moreover, the transversal noise coefficients lead to a frequency shift towards the DC ($0$~Hz) direction, which is not obvious at first glance as $\left| \vec{\mathbf B}\right| <\left| \vec{\mathbf B}+\delta \vec{\mathbf{B}}(t) \right|$ in case of a transversal noise. This becomes plausible when we recall that the transversal noises change the B-field direction so that the trajectory of the spin vector wobbles (in the longitudinal direction) during the precession. If this trajectory is projected onto the transversal plane, the resulting precession angle may observed to be smaller. This intuition is captured by an example in Fig.~\ref{fig:wobbling}, which describes a repeated rotation of $\vec{\mathbf{S}}(t)$ with respect to two different precession axes $\vec{\mathbf B}_{\text{tot}}= \vec{\mathbf B}+\delta \vec{\mathbf{B}}(t)$.

The intuition presented as in Fig.~\ref{fig:wobbling} is implicitly relying on the assumption that $\delta \vec{\mathbf{B}}(t)$ has a definite axis (here, the $x$ axis). As a result, one might suspect that such an anisotropy of noise might be important for the predicted Larmor frequency shift. Indeed, this is the case as indicated by the difference term $(\Gamma_{xx} - \Gamma_{yy})^2$ appearing in Eq.~\eqref{eq:omegasubd}. Moreover, as can be seen from Eq.~\eqref{eq:omegasubd}, if the transversally anisotropic noise is very large, $\omega_d$ will be an imaginary number and in this case the autocorrelation function is \textit{overdamped}; there will be no oscillations in the autocorrelation function.
The longitudinal-transversal correlations $\Gamma_{zx}$ and $\Gamma_{yz}$ do not seem to admit straightforward effective descriptions (however, see Appendix C).

We note that although the relaxation rates of the autocorrelation function (effective relaxation rates) are altered, the spin relaxation rates themselves are not altered by B-field noises; after all, once the B-field noise $\delta B(t)$ is given, the cross product term $\gamma \vec{\mathbf{S}} \times \delta \vec{\mathbf{B}}(t)$ is always orthogonal to $\vec{\mathbf{S}}$, it does not lead to an additional relaxation of the squared norm $\vec{\mathbf{S}} \cdot \vec{\mathbf{S}}$.

\section{Comparison between the theory and stochastic simulations} \label{sec:mc}
In this section, we numerically deal with cases in which the external noise is given as Gaussian white noise along the $x$ direction or the $y$ direction. We employ the Euler-Maruyama method, a direct generalization of the Euler method to the stochastic case, to numerically solve the corresponding stochastic differential equations. We refer to \cite{HAI2001} for a less technical review of the Euler-Maruyama method, section 10.2 of \cite{KNS1995} for a convergence proof of the method, and \cite{EIT2014, KSP2007} for an introduction to stochastic differential equations. In case $\delta \vec{\mathbf B} = \delta B \mathbf{\hat n}$ with $\left< \delta B(t) \delta B(t') \right> = \Gamma \delta(t-t')$, the Bloch equation \eqref{eq:bloch} can be re-written in the following form, which is convenient for computer simulations:
\begin{equation}
\begin{split}
d\vec{\mathbf S} &= \left( \begin{bmatrix} -\Gamma_2 & \omega &0 \\ -\omega & -\Gamma_2 & 0\\ 0 & 0&-\Gamma_1 \end{bmatrix}\vec{\mathbf S}+ \frac{R}{2} \mathbf{ \hat z} \right) dt\\
&+ \gamma \sqrt{\Gamma}\begin{bmatrix} 0&n_z&-n_y\\-n_z&0&n_x\\n_y&-n_x&0\end{bmatrix} \vec{\mathbf S} dW,
\end{split}
\end{equation}
where $W$ denotes the standard one-dimensional Wiener process (Brownian motion).

\begin{figure}[!h]
\begin{minipage}{\linewidth}
\centering
\subfloat[]{\label{fig:100PSD}\includegraphics[width=\linewidth]{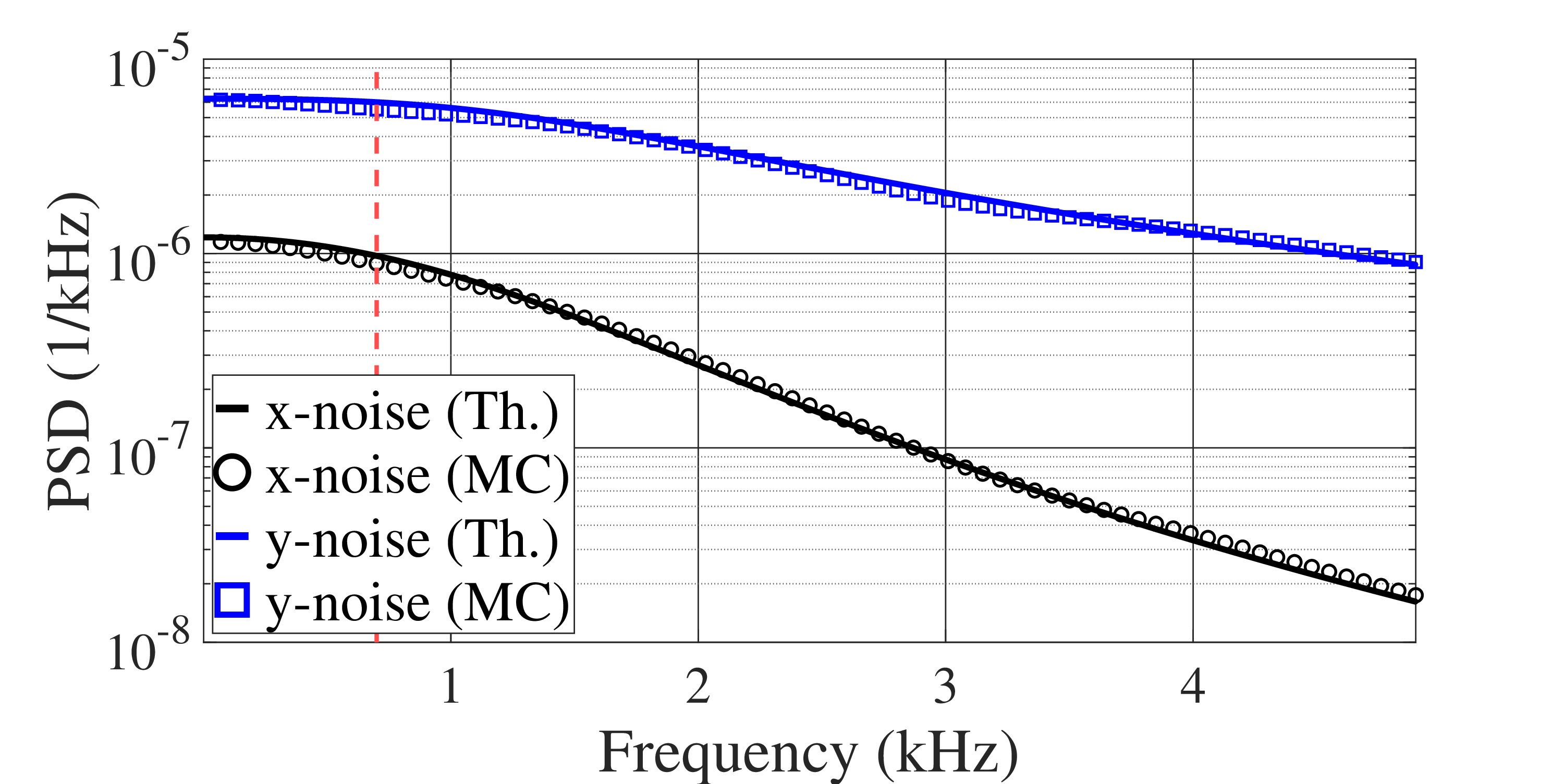}}
\end{minipage}\par\medskip
\begin{minipage}{\linewidth}
\centering
\subfloat[]{\label{fig:1000PSD}\includegraphics[width=\linewidth]{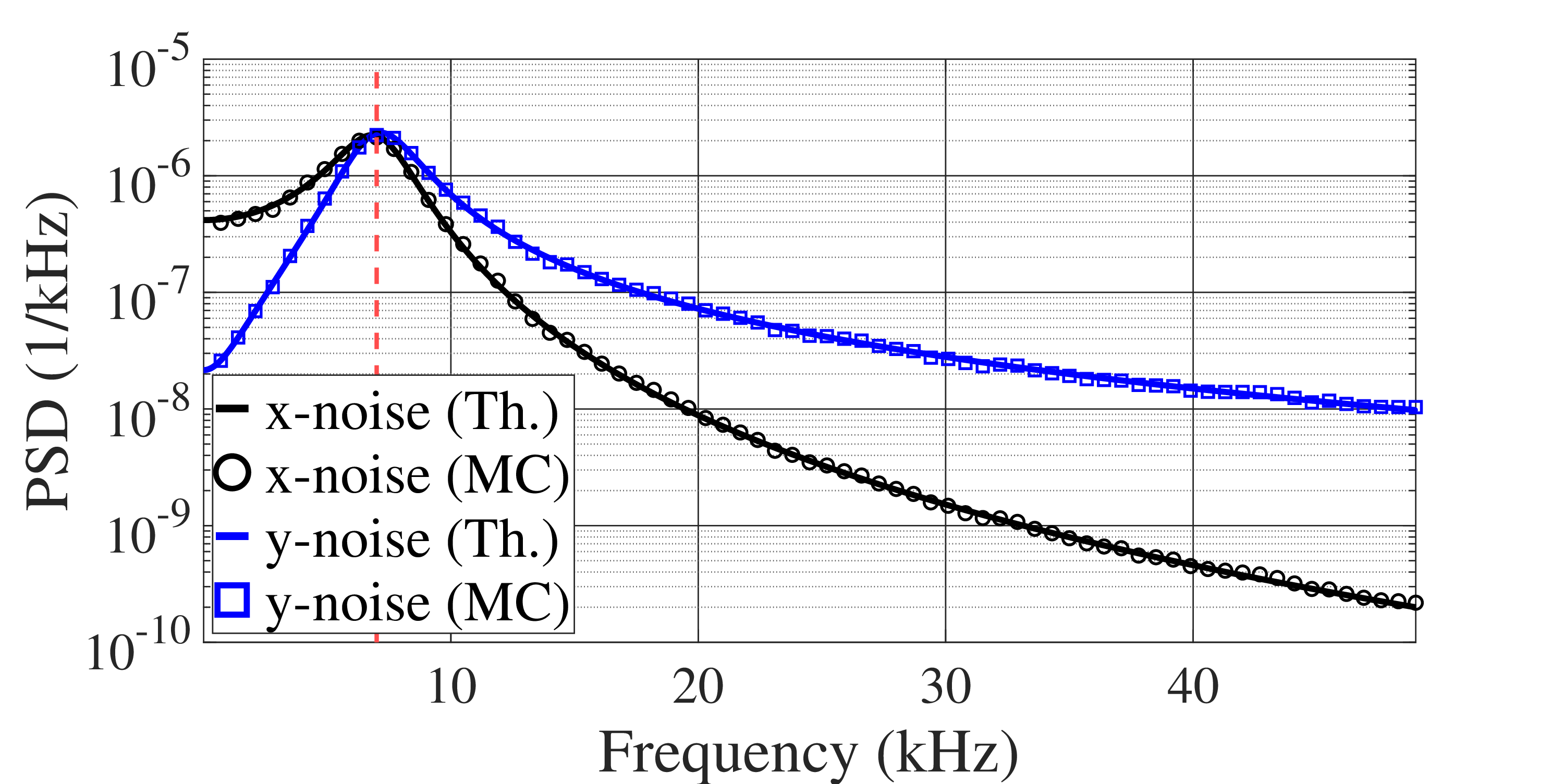}}
\end{minipage}
\caption{(a) Theory versus Monte Carlo where $B = 100$ nT. Black plots correspond to x-directional noise input whereas blue plots correspond to y-directional noise input. Monte Carlo simulations were repeated $100$ times and the averaged results are shown. The dashed red line corresponds to the Larmor frequency. The theory and the Monte Carlo simulations gave almost the same results. (b) This is for the case in which $B = 1000$ nT, with all other things being equal. We note that it tended to be much more difficult to deal with cases with stronger B-fields; the Larmor oscillations became quite rapid so that shorter time step sizes seemed to be required.}
\label{fig:PSD}
\end{figure}

Numerical analyses were performed with $R = 8\text{ ms}^{-1}$, $\Gamma_1 = 25/3 \text{ ms}^{-1}$, $\Gamma_2 = 10\text{ ms}^{-1}$, and $\gamma = 2\pi \times 0.007 \text{ rad/(ms} \cdot\text{nT)}$ (this $\gamma$ value corresponds to the case of $^{87}$Rb). The $z$-directional magnetic field was set to $100$ nT (Fig.~\ref{fig:100PSD}) or $1000$ nT (Fig.~\ref{fig:1000PSD}), and the noises were characterized by $\Gamma_{\alpha \alpha} = 1 \text{ nT}^2 \cdot \text{ms}$, where $\alpha=x$ or $y$; this corresponds to approximately $32$ pT$/\sqrt{\text{Hz}}$. The stochastic Bloch equation was integrated for a time period of $12$ ms, starting from its (zeroth-order) macroscopic equilibrium $\frac{R}{2\Gamma_1} \mathbf{\hat z}$. The last $6$ ms of the approximate solutions were used for further investigations, such as PSD calculations by Welch's method. The number of forward time steps was at least $2^{17}$ for each sample trajectory corresponding to a discretized Brownian motion (details are provided in Appendix B).

The results (Fig.~\ref{fig:PSD}) indicate that our analytic results are indeed correct. Note the relatively ``high-passing'' nature of the $y$-axis compared to the $x$-axis; this provides evidence to our intuition-based predictions.

\begin{figure}[!h]
\centering
\includegraphics[width=0.5\textwidth]{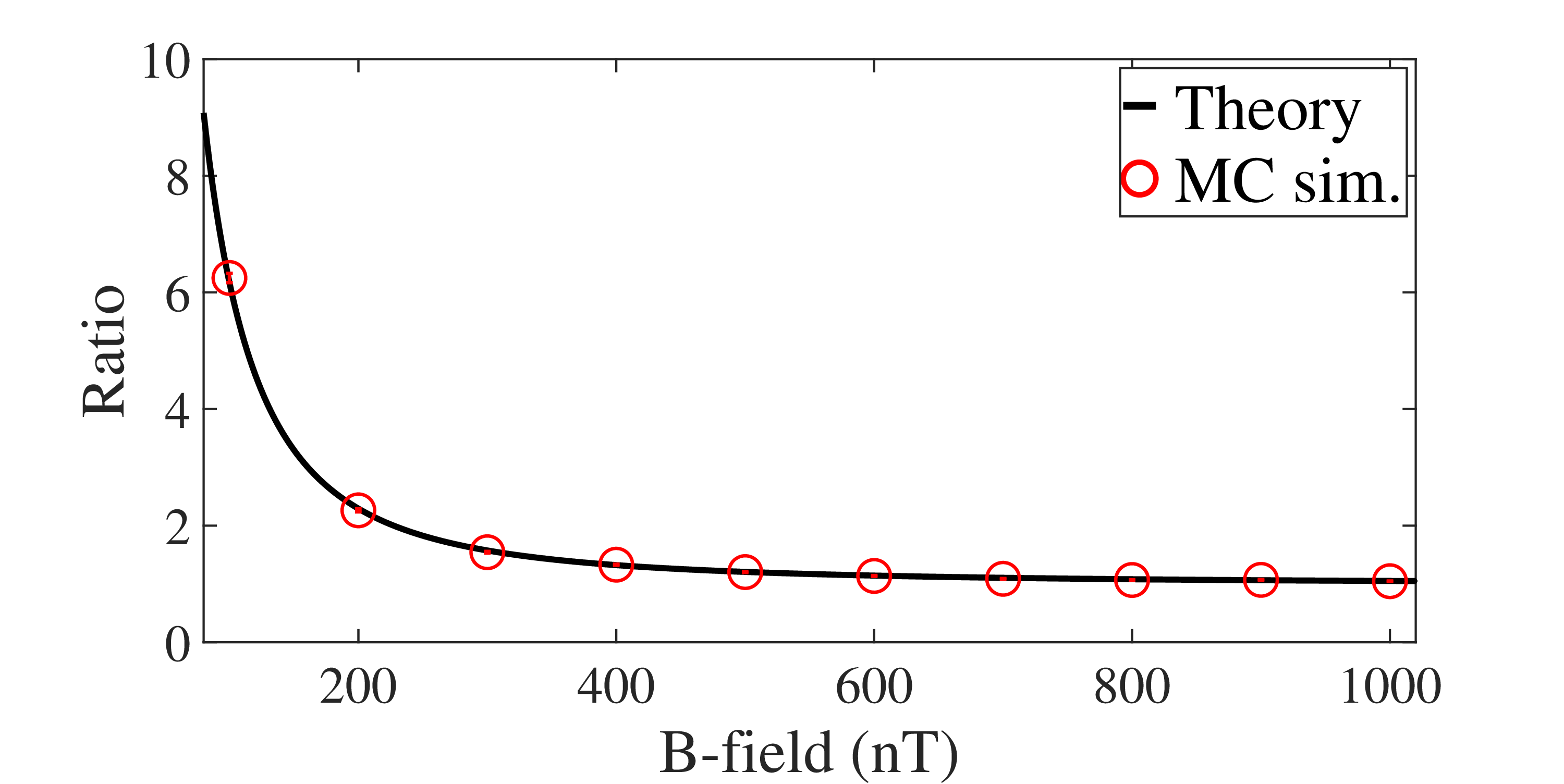}
\caption{Theory vs. Monte Carlo for the ratio described in Eq.~\eqref{eq:ratio}. Each point comes from an average of $50$ independent Monte Carlo experiments.}
\label{fig:ratio}
\end{figure}

Moreover, we plot (Fig.~\ref{fig:ratio}) the following ratio as a function of $B$, both in silico and in theory:
\begin{equation} \label{eq:ratio}
\frac{\text{PSD at resonance when noise is along y-axis}}{\text{PSD at resonance when noise is along x-axis}}.
\end{equation}
Fig.~\ref{fig:ratio} shows that the theory and the numerical experiment are consistent.

\begin{figure}[!h]
\centering
\includegraphics[width=0.5\textwidth]{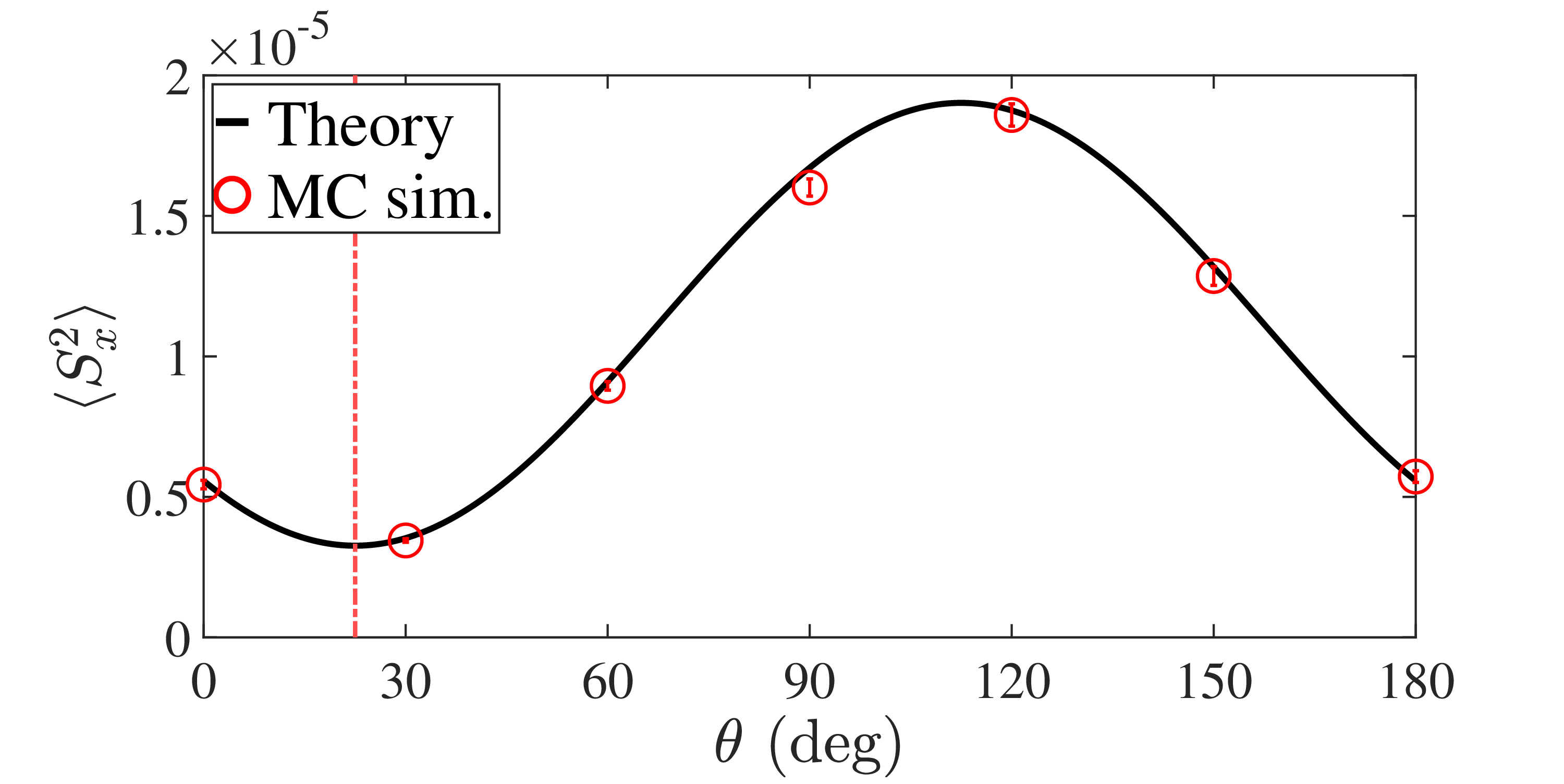}
\caption{Theory vs. Monte Carlo, for noise power described in Eq.~\eqref{eq:turnaround}. Here, the B-field is set to make $\omega = \Gamma_2$. Each point comes from an average of $50$ independent Monte Carlo experiments. The dash-dotted red line indicates for the theoretical minimum.}
\label{fig:turnaround}
\end{figure}

As a further illustration, we investigated the situation in which the noise direction is given by the unit vector $\mathbf{\hat n}=  \cos\theta  \mathbf{\hat x}+\sin\theta \mathbf{\hat y} $ by numerically verifying the proposed equation \eqref{eq:turnaround}. Now, the longitudinal B-field is set to $B=227.36$ nT (so that $\omega=\Gamma_2$) whereas the noise is characterized by $\Gamma_{\perp} = 1 \text{ nT}^2 \cdot \text{ms}$. The Monte Carlo results were obtained by directly calculating the time-average value of $S_x ^2$. The results are given in Fig.~\ref{fig:turnaround}, which show an agreement between the theory and the experiment.

\begin{figure}[t]
\centering
\includegraphics[width=0.5\textwidth]{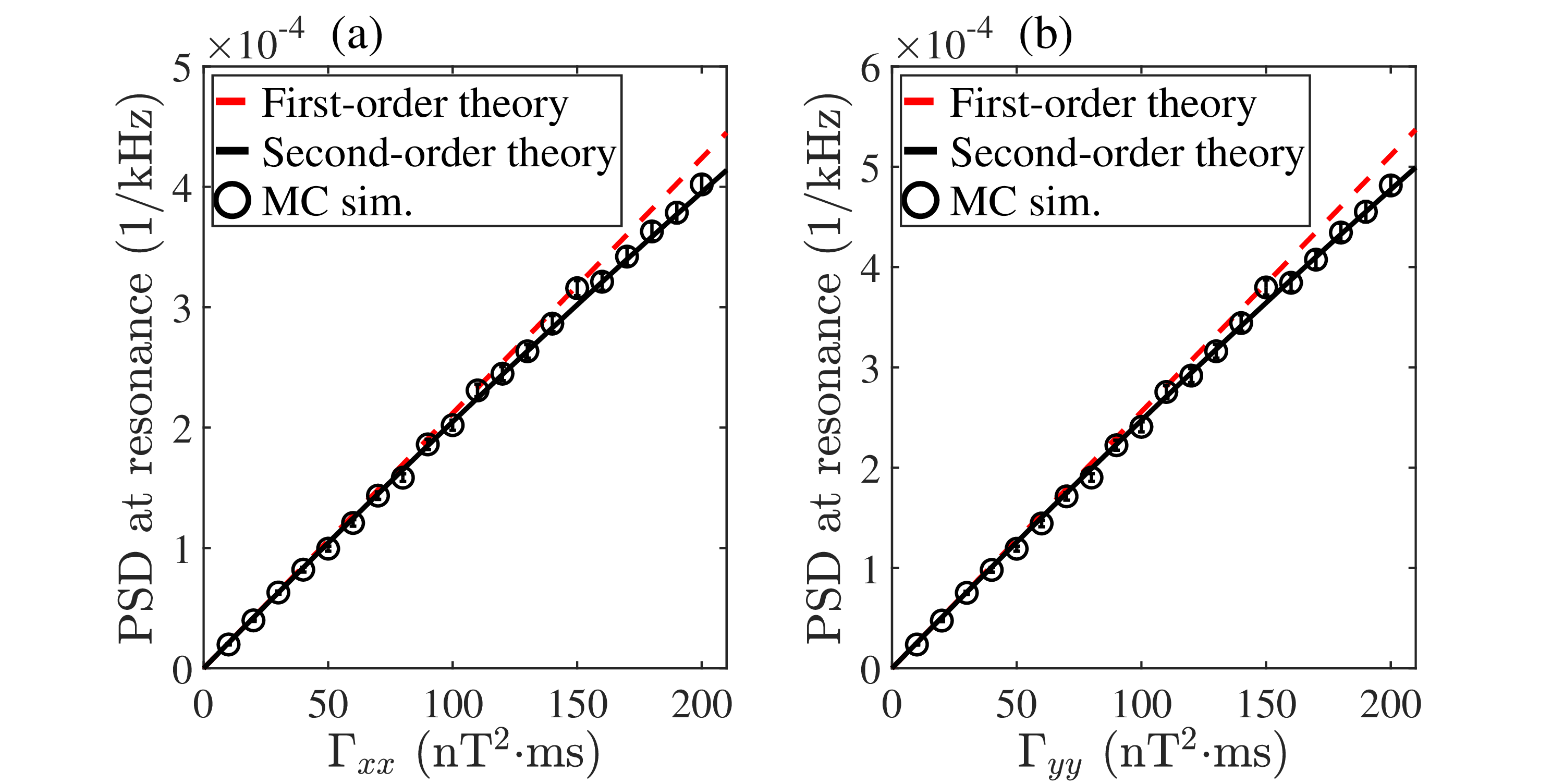}
\caption{Theory (up to first and second-order) versus Monte Carlo, for PSD values at resonance as a function of $\Gamma_{\alpha \alpha}$. Here, $B = 500$ nT. Subfigure (a) corresponds to x-directional noises, whereas subfigure (b) corresponds to y-directional noises. Each point comes from an average of $300$ independent Monte Carlo experiments.
}
\label{fig:FScom}
\end{figure}

Finally, the second-order results are supported by numerical calculations. With the same relaxation and pumping parameters as before, and with $B = 500$ nT, we plot the PSD value at resonance as a function of $\Gamma_{xx}$ or $\Gamma_{yy}$ (Fig.~\ref{fig:FScom}). For these calculations, we employed the first-order-corrected steady state value given in equation \eqref{eq:Szequil} as the initial condition. Note that the result shows that the second-order analysis is indeed more accurate.
We finally note that because both second-order coefficients at resonance are always negative (for non-zero $R$), the higher order terms should contribute positively to the PSD at resonance to prevent the value from being negative.

\section{Conclusions}
We have solved, by a time-dependent perturbative approach, the optically pumped stochastic Bloch equation describing a macroscopic spin collection subjected to small magnetic field white noises in the presence of an external bias magnetic field and a $\sigma^+$ pumping beam, both along the $z$ direction. We have focused on the regime where the magnetic field noise-driven spin precession dominates the effect of the intrinsic spin noise so that the latter can be considered negligible. The analytic forms for the autocorrelation function and the corresponding power spectral density (PSD) of the macroscopic spin along the probing axis ($x$-axis in our case) are given up to first-order. The analytic formula of the first-order PSD is of a non-Lorentzian nature. Moreover, it reveals anisotropy in the effects of the magnetic field noises; the effect of $x$-directional magnetic field noises dominates at the low frequency range of the PSD (given $\omega \gg \Gamma_2$) whereas the $y$-directional magnetic field noises prevail at high frequencies $\Omega \gg \omega$. The $z$-directional magnetic field noises, represented by $\Gamma_{zz}$, $\Gamma_{zx}$, and $\Gamma_{yz}$, contribute to the PSD via second-order corrections. A closer inspection shows that anisotropic transversal B-field noises will lead to an effective Larmor frequency shift towards the DC direction when one observes the PSD. Our analytic results are supported by Monte Carlo simulations approximately solving the stochastic Bloch equation. Our analysis will be useful for understanding the behavior of a spin collection in a noisy magnetic field, which is a situation naturally arising in studies of atomic magnetometers.

\appendix
\section{Simulation parameters for Fig.~\ref{fig:highlowexp}}
For Fig.~\ref{fig:highlowexp}, the simulation was again done using the Euler-Maruyama method. $B$ was set to $500 \text{ nT}$; other parameters were the same as in Section \ref{sec:mc}. The input noises were simulated by applying high-pass and low-pass filters to a prescribed transversal Gaussian white noise. 
These filtering processes were done by MATLAB R2022a \texttt{highpass} and \texttt{lowpass} functions with their cutoff frequencies being 10 times the Larmor frequency and 0.1 times the Larmor frequency, respectively.

The white noise $\delta \vec {\mathbf B} = \delta B \mathbf{\hat n}$, where $\mathbf{\hat n} \perp \mathbf{\hat{z}}$, satisfied $\langle \delta B (t) \delta B(t') \rangle = (1 \text{ nT}^2 \cdot \text{ms}) \text{ }\delta(t-t')$. Indeed, noise-parallel spin magnitudes were smaller than the noise-perpendicular spin magnitudes when the input noise consisted of high-frequency components.

\section{Determination of time step size for Euler-Maruyama method}
The time step sizes for the numerical simulations were determined by a relative error criterion. First, the whole time interval $[0,12]$ (ms) was divided into $2^{25}$ subintervals of equal lengths and a discretized Brownian motion was generated accordingly. Then, we executed the Euler-Maruyama method with time step size $\Delta t = 12/2^{16}$. With such a time step size, each Brownian increment $\Delta W$ for a single time step of the method was obtained by summing $2^{25-16}=512$ consecutive entries of the discretized Brownian motion. Then, we reduced the time step size by a factor of $1/2$, and the whole Euler-Maruyama method was repeated; such half-reductions of the time step were performed until the relative errors between the previous calculation and the current calculation were always less than $0.01$ at time instances $t = 12-i\times12/2^6 $ for $i=0,1,\cdots,29$. This condition was checked separately for all three components of the spin vector (so that $30 \times 3 = 90$ relative error conditions were checked in total). If the condition failed until $\Delta t$ reached $12/2^{25}$, the corresponding discretized Brownian motion was discarded since it led to failure.
The above-described procedure was also used for determining the time step size for Fig.~\ref{fig:highlowexp}; after the time step size was determined, corresponding $\Delta W$ sequence was calculated and then fed into the high-pass and low-pass filters described in Appendix A.

\section{Omitted formulae}
\begin{strip}
$\mathfrak{C}$ and $\mathfrak{D}$, appeared in Eq.~\eqref{eq:originalcoordinates}, are given as
\begin{equation} \label{eq:frakC}
\mathfrak{C}=\gamma^2 \frac{ \left( e^{-\Gamma_1 \tau}- e^{-\Gamma_2 \tau} \cos(\omega\tau)\right) \left(\omega \Gamma_{yz} - \Gamma_1 \Gamma_{zx} + \Gamma_2 \Gamma_{zx}   \right)+e^{-\Gamma_2 \tau} \left(\Gamma_1 \Gamma_{yz} - \Gamma_2 \Gamma_{yz} + \omega \Gamma_{zx}\right) \sin(\omega\tau)} {2 \left( \omega^2 + \left(\Gamma_1 - \Gamma_2 \right)^2 \right) }
\end{equation}
and
\begin{equation} \label{eq:frakD}
\begin{split}
\mathfrak{D}&=\frac{\gamma^2 R}{4 \Gamma_1 \left( \omega^2 + \Gamma_2 ^2 \right) \left( \omega^2 + \left(\Gamma_1 - \Gamma_2\right)^2 \right)} \left\{ \vphantom{\int_0^0} \left( \omega \Gamma_{yz} +\Gamma_2 \Gamma_{zx} \right)  \left( \omega^2 + \left(\Gamma_1 - \Gamma_2 \right)^2 \right) \right. \\
&- e^{-\Gamma_1 \tau}  \left( \omega^2 +\Gamma_2 ^2 \right) \left( \omega \Gamma_{yz}  -\Gamma_1 \Gamma_{zx} + \Gamma_2 \Gamma_{zx} \right)\\
&- e^{-\Gamma_2 \tau} \Gamma_1 \left(\Gamma_1 \Gamma_2 \Gamma_{zx} - \Gamma_2 ^2 \Gamma_{zx} + \Gamma_1 \Gamma_{yz} \omega - 2 \Gamma_2 \omega \Gamma_{yz} + \omega^2 \Gamma_{zx} \right) \cos(\omega \tau)\\
&\left. - e^{-\Gamma_2 \tau} \Gamma_1 \left(\Gamma_1 \Gamma_2 \Gamma_{yz} - \Gamma_2 ^2 \Gamma_{yz} - \Gamma_1 \Gamma_{zx} \omega + 2 \Gamma_2 \omega \Gamma_{zx} + \omega^2 \Gamma_{yz} \right) \sin(\omega \tau)\vphantom{\int_0^0} \right\},
\end{split}
\end{equation}
respectively.
Somewhat interestingly, $\mathfrak{D} \simeq \frac{\gamma^2 R \left( \omega \Gamma_{yz} + \Gamma_2 \Gamma_{zx} \right)}{4 \Gamma_1 \left( \omega^2 + \Gamma_2 ^2 \right)}$ for $\Gamma_1 \tau \text{, } \Gamma_2 \tau \gg 1$, and this expression equals the right-hand-side of Eq.~\eqref{eq:Sxequil}.
Also, note that the equations~\eqref{eq:frakC} and \eqref{eq:frakD} exhibit some longitudinal-transversal coupling terms involving $\Gamma_{zx}$ or $\Gamma_{yz}$ which are not multiplied by a sinusoidal term. Therefore, if these coupling terms are Fourier transformed to give the PSD of $S_x$, this will result in a DC noise component centered at $0$~Hz, which is somewhat qualitatively similar to the double peak phenomenon noted in \cite{PSN2014}. Indeed, if one performs the relevant second-order calculations, the following second-order DC term occurs in the autocorrelation function of $S_x$:
\begin{equation}\frac{\gamma^4 R^2}{16 \Gamma_1^2 \left( \Gamma_2^2 +\omega^2 \right)^2} \left( \Gamma_2 \Gamma_{zx} + \omega \Gamma_{yz} \right)^2.
\end{equation}

$\mathbf{\widetilde {\Gamma}}$, appeared in Eq.~\eqref{eq:approxODE}, is given as:
\begin{equation}
\mathbf{\widetilde {\Gamma}} = \gamma^2 \begin{bmatrix} \mathfrak{A} & \mathbf{0} & \mathbf{0}\\ \mathbf{0} &\mathfrak{B}& \mathbf{0}\\ \mathbf{0}&\mathbf{0}&0\end{bmatrix},
\end{equation}
where \begin{equation}
 \mathfrak{A} = \begin{bmatrix}
 -b-c & \frac{d}{2} & \frac{f}{2} & \frac{d}{2} & c & -e & \frac{f}{2} & -e & b \\[0.2cm]
 \frac{d}{2} & -\frac{a}{2}-\frac{b}{2}-c & \frac{e}{2} & -c & \frac{d}{2} & f & e & \frac{f}{2} & -d \\[0.2cm]
 \frac{f}{2} & \frac{e}{2} & -\frac{a}{2}-b-\frac{c}{2} & e & -f & \frac{d}{2} & -b & d & \frac{f}{2} \\[0.2cm]
 \frac{d}{2} & -c & e & -\frac{a}{2}-\frac{b}{2}-c & \frac{d}{2} & \frac{f}{2} & \frac{e}{2} & f & -d \\[0.2cm]
 c & \frac{d}{2} & -f & \frac{d}{2} & -a-c & \frac{e}{2} & -f & \frac{e}{2} & a \\[0.2cm]
 -e & f & \frac{d}{2} & \frac{f}{2} & \frac{e}{2} & -a-\frac{b}{2}-\frac{c}{2} & d & -a & \frac{e}{2} \\[0.2cm]
 \frac{f}{2} & e & -b & \frac{e}{2} & -f & d & -\frac{a}{2}-b-\frac{c}{2} & \frac{d}{2} & \frac{f}{2} \\[0.2cm]
 -e & \frac{f}{2} & d & f & \frac{e}{2} & -a & \frac{d}{2} & -a-\frac{b}{2}-\frac{c}{2} & \frac{e}{2} \\[0.2cm]
 b & -d & \frac{f}{2} & -d & a & \frac{e}{2} & \frac{f}{2} & \frac{e}{2} & -a-b
\end{bmatrix}
\end{equation}
and \begin{equation}
 \mathfrak{B} = \begin{bmatrix}
-\frac{b}{2}-\frac{c}{2} & \frac{d}{2} & \frac{f}{2} \\[0.2cm]
 \frac{d}{2} & -\frac{a}{2}-\frac{c}{2} & \frac{e}{2} \\[0.2cm]
 \frac{f}{2} & \frac{e}{2} & -\frac{a}{2}-\frac{b}{2}
\end{bmatrix},
\end{equation}
wherein we defined $a\coloneqq\Gamma_{xx}$, $b\coloneqq\Gamma_{yy}$, $c\coloneqq\Gamma_{zz}$, $d\coloneqq\Gamma_{xy}$, $e\coloneqq\Gamma_{yz}$, and $f\coloneqq\Gamma_{zx}$ to save space.
Also, in Eq.~\eqref{eq:secondordertermDK},
\begin{equation}
\mathcal{D} = \begin{bmatrix}
-\Gamma_{yy}-\Gamma_{zz} & \Gamma_{xy} & \Gamma_{zx} &0\\
\Gamma_{xy} & -\Gamma_{zz}-\Gamma_{xx}&\Gamma_{yz}&0\\
\Gamma_{zx}&\Gamma_{yz} & -\Gamma_{xx}-\Gamma_{yy}&0\\
0&0&0&0
\end{bmatrix}.
\end{equation}

Finally, we provide the second-order corrections for the PSD of $S_x$ corresponding to the products $\Gamma_{xx}\Gamma_{zz}$, $\Gamma_{yy}\Gamma_{zz}$, and $\Gamma_{xx}\Gamma_{yy}$. They are given as 
\begin{equation}
\frac{\gamma^4 R^2  \left( \mathcal C_1 \Gamma_{xx}\Gamma_{zz} +\mathcal C_2 \Gamma_{yy}\Gamma_{zz}\right) }{8 \Gamma_1^2 \Gamma_2 \left(\omega^2 + \Gamma_2^2\right)  \left( (\Omega-\omega)^2+\Gamma_2^2 \right)^2 \left( (\Omega+\omega)^2+\Gamma_2^2 \right)^2}
\end{equation}
and
\begin{equation}
\frac{-\gamma^4 R^2   \mathcal C_3 \Gamma_{xx}\Gamma_{yy} }{2 \Gamma_1^3 \left( (\Omega-\omega)^2+\Gamma_2^2 \right)^2 \left( (\Omega+\omega)^2+\Gamma_2^2 \right)^2}
\end{equation}
where $\mathcal C_i$ are given by the following equations.
\begin{subequations}
\begin{align}
\mathcal C_1 &= \left( \omega^2 + 2 \Gamma_2 ^2 \right) \Omega^6 - \left( \omega^4 +3\Gamma_2 ^2 \omega^2 -6\Gamma_2^4 \right) \Omega^4 - \left(\omega^6 + 9 \Gamma_2^4 \omega^2 - 6 \Gamma_2^6  \right)\Omega^2 + \left( \omega^8- 7 \Gamma_2^2 \omega^6 - 15\Gamma_2^4 \omega^4 -5 \Gamma_2^6 \omega^2 +2\Gamma_2^8 \right)\\
\mathcal C_2 &=  \omega^2 \Omega^6 - \left( \omega^4 -3\Gamma_2 ^2 \omega^2 +4\Gamma_2^4 \right) \Omega^4 - \left(\omega^6 + 22 \Gamma_2^2 \omega^4 +13 \Gamma_2^4 \omega^2 +8\Gamma_2^6  \right)\Omega^2 + \left( \omega^8+11 \Gamma_2^2 \omega^6 + 15\Gamma_2^4 \omega^4 + \Gamma_2^6 \omega^2 -4\Gamma_2^8 \right)\\
\mathcal C_3 &=  \Omega^6 + \left( 3\Gamma_2^2 -\omega^2 \right) \Omega^4 - \left(\omega^4 - 2 \Gamma_2 (2\Gamma_1 +\Gamma_2) \omega^2 -3\Gamma_2^4 \right)\Omega^2 + \left( \omega^2 + \Gamma_2^2 \right)^3
\end{align}
\end{subequations}

\end{strip}

\section*{Acknowledgments}
The authors thank Ji Hoon Yoon for having a fruitful discussion with the authors about the topic. This work was supported by the Agency for Defense Development Grant funded by the Korean government (915001102).

\end{document}